\documentclass[journal]{IEEEtran}

\usepackage{amsmath}
\usepackage{amsthm}
\usepackage{amsfonts}
\usepackage{amssymb}
\usepackage{epsfig}
\usepackage{psfrag}
\usepackage{graphicx}
\usepackage{float}
\usepackage{multirow}
\usepackage{tabularx}
\usepackage{color}
\usepackage{algorithm}
\usepackage{algorithmic}
\usepackage{lineno}
\usepackage{hyperref}
\usepackage{multicol}
\usepackage{commath}
\usepackage{stfloats}
\usepackage{wrapfig}
\usepackage{subcaption}
\usepackage{caption}
\usepackage{cleveref}
\usepackage{tikz}
\usetikzlibrary{shapes.geometric, arrows}
\usepackage{adjustbox}
\usepackage{pifont}
\usepackage[sort&compress,square,sort,comma,numbers]{natbib}
\usepackage{soul}

\newcolumntype{C}[1]{>{\centering\let\newline\\\arraybackslash\hspace{-2mm}}m{#1}}

\tikzstyle{startstop} = [rectangle, rounded corners, minimum width=2cm, minimum height=1cm, text centered, draw=black, fill=white]
\tikzstyle{io} = [trapezium, trapezium left angle=70, trapezium right angle=110, minimum width=2cm, minimum height=1cm, text width=4cm, text centered, draw=black, fill=white]
\tikzstyle{process} = [rectangle, minimum width=2cm, minimum height=1cm, text width=3cm, text centered, draw=black, fill=white]
\tikzstyle{decision} = [diamond, minimum width=2cm, minimum height=1cm, text width=2cm, text centered, draw=black, fill=white]
\tikzstyle{arrow} = [thick,->,>=stealth]

\begin{document}

\date{}

\title{Voting Scheme to Strengthen Localization Security in Randomly Deployed Wireless Sensor Networks}

\author{Slavisa~Tomic,
        Marko~Beko,
        Dejan Vukobratovic,
        and Srdjan Krco
\thanks{This research was partially funded by the European Union’s Horizon Europe Research and Innovation Programme under the Marie Sk\l{}odowska-Curie grant agreement No. 101086387 and by Funda\c{c}\~{a}o para a Ci\^{e}ncia e a Tecnologia under Projects UIDB/04111/2020, UIDB/50008/2020, ROBUST EXPL/EEI-EEE/0776/2021, and 2021.04180.CEECIND.}
\thanks{S.~Tomic is with Universidade Lus\'{o}fona, Campo Grande 376, 1749-024 Lisboa, Portugal. (e-mail: slavisa.tomic@ulusofona.pt). M.~Beko is with Instituto de Telecomunica\c{c}\~{o}es, Instituto Superior T\'{e}cnico, Universidade de Lisboa, 1049-001 Lisbon, Portugal (e-mail: marko.beko@tecnico.ulisboa.pt). S.T. and M.B. are also with Center of Technology and Systems (UNINOVA-CTS) and Associated Lab of Intelligent Systems (LASI),  2829-516 Caparica, Portugal. D.~Vukobratovic is with Faculty of Technical Sciences, University of Novi Sad, 21000 Novi Sad, Serbia (e-mail: dejanv@uns.ac.rs). S.~Krco is with DunavNET, Bul. Oslobodjenja 133/2, 21000 Novi Sad, Serbia (e-mail: srdjan.krco@dunavnet.eu).}}

\maketitle



\begin{abstract}
This work aspires to provide a trustworthy solution for target localization in adverse environments, where malicious nodes, capable of manipulating distance measurements (i.e., performing spoofing attacks), are present, thus hindering accurate localization. Besides localization, its other goal is to identify (detect) which of the nodes participating in the process are malicious. This problem becomes extremely important with the forthcoming expansion of IoT and smart cities applications, that depend on accurate localization, and the presence of malicious attackers can represent serious security threats if not taken into consideration. This is the case with most existing localization systems which makes them highly vulnerable to spoofing attacks. In addition, existing methods that are intended for adversarial settings consider very specific settings or require additional knowledge about the system model, making them only partially secure. Therefore, this work proposes a novel voting scheme based on clustering and weighted central mass to securely solve the localization problem and detect attackers. The proposed solution has two main phases: 1) Choosing a cluster of suitable points of interest by taking advantage of the problem geometry to assigning votes in order to localize the target, and 2) Attacker detection by exploiting the location estimate and basic statistics. The proposed method is assessed in terms of localization accuracy, success in attacker detection, and computational complexity in different settings. Computer simulations and real-world experiments corroborate the effectiveness of the proposed scheme compared to state-of-the-art methods, showing that it can accomplish an error reduction of $30~\%$ and is capable of achieving almost perfect attacker detection rate when the ratio between attacker intensity and noise standard deviation is significant.
\end{abstract}

\begin{IEEEkeywords}
Secure localization, voting scheme (VS), clustering, received signal strength (RSS), weighted central mass (WCM), wireless sensor networks (WSNs), spoofing attacks.
\end{IEEEkeywords}



\section{Introduction}
\label{sec:intro}

Recently, wireless sensor networks (WSNs) have attracted much interest of the scientific community, partially due to their ability to work in harsh environments, ease and low costs of implementation~\cite{Tomic:2021, Ricardo:2021}, and wide variety of applications~\cite{Oigbochie:2021, cheng:2022}. From the localization perspective, generally, WSNs are composed of two distinct types of nodes: 1) anchor nodes, whose locations are known and serve as reference points in the localization process and 2) target nodes, whose locations are unknown and one desires to determine them. Naturally, it is expected that nodes are capable to communicate with each other in order to execute the localization task. In this work, a non-cooperative network, where targets are only allowed to communicate with anchors, is considered.

In most applications, data acquired by sensors are only useful if they can be associated with the respective physical location. However, most existing localization systems overlook possible security threats~\cite{TomicBeko:2020, Coluccia:2018}. Therefore, if these systems are exposed to malicious attacks, they can result in catastrophic outcomes (e.g., failure in a self-driving car collision system, change in drone trajectory, etc.). Mainly for this reason, localization systems should be developed for potentially adversarial environments, where a malicious (or damaged) sensor can produce false distance measurements (spoofing attacks) to manipulate the localization process.

Perhaps the easiest and most common way of localization is to equip sensors with global positioning system (GPS) receivers. However, this solution has several undesirable consequences, such as increased implementation costs and infeasibility in some environments (e.g., indoor, urban areas, forests, etc.). In addition, from the security point of view, GPS is considered a civilian localization system (e.g., uses unencrypted signals); thus, it is not very difficult to manipulate (spoof) it. Although there are different types of attacks in WSNs nowadays, this work focuses its attention on spoofing attacks specifically due to their close relationship with distance measurements (proximity). These attacks include forging distance measurements (either by reducing or enlarging them) that can be performed in various non-cryptographic ways, meaning that no infraction of upper-layer security protocols for carrying out these attacks is at risk (for instance, an attacker can add a physical obstacle between two nodes or change node's transmit power without informing its neighbours). Moreover, these attacks (including GPS spoofing) come at a relatively low cost, but can cause severe problems in many systems~\cite{Kugler:2017}. Since spoofing attacks are closely related with distance they have real-world impact and are used in car thefts, executions of unauthorized payments and manipulation of navigation~\cite{Singh:2019}.

\subsection{Related Work}
\label{subsec:related_work}

Secure localization in WSNs has attracted interest in the scientific literature~\cite{Li:2005, Liu:2007, Garg:2012, Liu:2019, Li:2021, Beko:2021, TomicS:2022, Mukhopadhyay:2021, Tomic_TVT1:2024, Tomic_TVT2:2024}. Still, there is no uniquely accepted solution and there is room for improvement in all aspects (localization accuracy, detection rates and complexity).

In~\cite{Li:2005}, two secure localization solutions were explored: least median of squares (LMS) and radio frequency (RF) fingerprinting. LMS selects the subset of anchors with the least median residues, while RF fingerprinting employs a median-based distance metric. The work in~\cite{Liu:2007} introduced attack-resistant minimum mean square estimation (ARMMSE) and a voting scheme (VS). ARMMSE detects and removes malicious anchors based on inconsistencies, whereas VS assigns votes to grid cells based on distance measurements, identifying the target's likely location. An iterative algorithm using gradient descent was proposed in~\cite{Garg:2012} for both uncoordinated and coordinated attacks. Malicious anchors were identified by their higher residues and excluded from the localization process. In~\cite{Liu:2019}, a density-based spatial clustering method classified location points as normal or abnormal, followed by a sequential probability ratio test to authenticate anchors using received signal strength (RSS) and time of arrival (TOA) data. The study in~\cite{Li:2021} addressed secure localization and velocity estimation in mobile WSNs with malicious anchors, using a maximum a posteriori (MAP) estimator solved via variational message passing. In~\cite{Beko:2021}, an initial location estimate was obtained via weighted central mass (WCM), followed by distance-based filtering and a generalized trust region sub-problem (GTRS) solved using a bisection method. This was extended in~\cite{TomicS:2022} to a general range-based scenario using a generalized likelihood ratio test (GLRT) and law of cosines (LC). The work in~\cite{Mukhopadhyay:2021} proposed secure weighted least squares (SWLS) for uncoordinated spoofing attacks and l1-norm (LN-1E) for coordinated attacks. SWLS filters malicious nodes based on estimated noise standard deviation, while LN-1E uses 3D plane fitting and K-means clustering to separate malicious anchors. In~\cite{Tomic_TVT1:2024}, a robust min-max approach was formulated as a second-order cone programming (R-SOCP) problem and a robust GTRS (R-GTRS). Lastly,~\cite{Tomic_TVT2:2024} introduced an alternating direction method of multipliers (ADMM) approach, applying a weighted least squares criterion within a decomposition-coordination iterative scheme.

Even though the methods in~\cite{Li:2005}-\cite{Tomic_TVT2:2024} work well in the settings under scrutiny in the respective works, they either require additional knowledge on certain parameters, such as noise power or the maximum magnitude of an attack (e.g.,~\cite{Liu:2007},~\cite{Liu:2019},~\cite{Li:2021},~\cite{Mukhopadhyay:2021},~\cite{Tomic_TVT1:2024}) and/or convex relaxations/approximations that expand the set of possible solution leading to higher error (e.g.,~\cite{Li:2005},~\cite{Li:2021}-\cite{Tomic_TVT1:2024}) or are executed iteratively (e.g.,~\cite{Garg:2012},~\cite{Tomic_TVT2:2024}), leaving room for improvement in all aspects (localization accuracy, detection rates and complexity). These limitations could severely deteriorate their performance in scenarios where the required parameters are unknown or imperfectly known, i.e., where problem assumptions do not hold, resulting in the applied relaxations/approximations not being sufficiently tight or lead to burdensome computations that might even raise convergence issues. Furthermore, the majority of the existing works assume knowledge about the type of spoofing attacks (uncoordinated or coordinated) under which the network is beforehand and develop a solution for that specific type of attack. However, it is not possible to acquire such knowledge in practice. Therefore, there are still some challenges to address and room for improvement in all main aspects of the considered problem.

\subsection{Main Contributions of the Current Work}
\label{subsec:contributions}

This work presents a geometric approach to compute intersection points between pairs of anchors and obtain an estimate of the target's location through a VS and WCM, which is then exploited to detect attackers based on confidence intervals. Unlike~\cite{Liu:2007,Garg:2012,Beko:2021,TomicS:2022,Mukhopadhyay:2021}, the proposed algorithm does not make hard (binary) detection decisions, but assigns beliefs (votes) to each intersection point, leveraging information from malicious anchors when attack intensity is low. Besides, mistakenly removing a genuine anchor can severely degrade the localization accuracy, leading to severe and potentially lethal consequences in real life, such as collisions of autonomous vehicles with obstacles, another vehicles or infrastructure, tardy arrival at the desired location in the case of an emergency event, like wildfire and organ transplantation and similar. 
The highest-voted points are converted into probabilities and used as WCM weights for location estimation. Attack intensities are assessed per link using a maximum likelihood (ML) criterion, with attacker detection based on confidence intervals at a predefined level. Due to its geometric nature, the method is adaptable to any range-based measurement. The primary goal is to advance secure localization beyond traditional systems, ensuring reliable malicious node detection and accurate target positioning. The main contributions of this work are threefold and are summarized in the following:
\begin{itemize}
\item{Design of a novel solution for target localization in randomly deployed sensor network in the presence of (uncoordinated and coordinated) spoofing attacks based on a new VS. The proposed localization estimator is a single-iteration scheme based on simple geometry, where points of interest extracted from the system model are clustered together and appraised by assigning votes to reach a set of the most trustworthy points. This set of points is then used to get an estimate on the target's location through WCM.}
\item{Unlike the existing methods that make a \emph{hard} binary decision, the current work converts votes into probabilities and makes a \emph{soft} detection decision, i.e., it makes use of all anchors and gradually builds its confidence regarding which anchors might be malicious and which not, not excluding any anchor in the process.}
\item{Proposal of a novel attacker detection scheme based on confidence intervals. By exploiting the location estimate, estimates of the attack intensities and the genuine (non-malicious) measurement values are acquired and used to evaluate if a reported measurement falls outside a confidence interval of one (estimated) standard deviation from its \textit{genuine} doublet.}
\item{The proposed solution not only matches, but outperforms more complex state-of-the-art solutions for high attack intensities. Thus, the proposed method sets the best achievable results reported in the literature thus far, both in terms of localization and detection performance.}
\item{Lastly, both computer simulations and real-world experimental measurements are employed to validate the performance of the proposed solution.}
\end{itemize}

Throughout the work, uppercase bold type, lowercase bold type, and regular type of fonts are used to denote matrices, vectors, and scalars, respectively. $\mathbb{R}^n$ denotes the $n$-dimensional real Euclidean space. Symbol $(\bullet)^T$ represent the transpose operator, while $\binom{n}{k} = \frac{n!}{k!(n-k)!}$ and $proj_{S}(\boldsymbol{p})$ denote the binomial coefficients and the projection of the point $\boldsymbol{p}$ onto the set $S$, respectively. The normal (Gaussian) distribution with mean $\mu$ and variance $\sigma^2$ is denoted by $\mathcal{N}(\mu, \sigma^2)$. The $N$-dimensional identity matrix is denoted by $\boldsymbol{I}_N$ and the $M \times N$ matrix of all zeros by $\boldsymbol{0}_{M \times N}$ (if no ambiguity can occur, subscripts are omitted). $\|\boldsymbol{x}\|$ denotes the vector norm defined by $\|\boldsymbol{x}\| = \sqrt{\boldsymbol{x}^T \boldsymbol{x}}$, where $\boldsymbol{x} \in \mathbb{R}^n$ is a column vector. The $i$-th column of the matrix $\boldsymbol{M}$ is denoted by $\boldsymbol{M}_i$ and an estimate of a parameter $\boldsymbol{x}$ with $\widehat{\boldsymbol{x}}$.

The rest of this paper is structured as follows: Section~\ref{sec:problem_formulation} defines the problem, while Section~\ref{sec:proposed_method} details the proposed solution. Section~\ref{sec:performance} presents performance comparisons, and Section~\ref{sec:conclusions} summarizes the key findings.



\section{Problem Formulation}
\label{sec:problem_formulation}

Let us consider a 2-dimensional, non-cooperative and static WSN, where a single target, whose true location is unknown and denoted by $\boldsymbol{x}$, is located at a time by the help of a set of anchors whose true locations are known and denoted by $\boldsymbol{a}_i, i=1,...,N$. Some of the anchors are assumed malicious and try to disrupt the location process by manipulating their distance measurements (spoofing attacks). The target receives radio signals (from which it measures the RSS values) from the anchors. Note that in this work the measurement acquisition is assumed practically instantaneous; thus, ambient conditions do not change significantly during the measurement process, having very little to no impact on the deviation of the RSS measurements. This work considers two types of spoofing attacks: coordinated and uncoordinated.

\subsection{Uncoordinated Attack}
\label{subsec:uncoord_att}

In this setting, the genuine anchors have a predefined transmitted power, while the malicious anchors change the transmitted power arbitrarily without notifying the network. The $k$-th RSS measurement sample ($k=1,...,K$) between the target and the $i$-th anchor can be modeled as
\begin{equation}
 P_{i,k} = P_0 - 10 \gamma \log_{10}\frac{\|\boldsymbol{x}-\boldsymbol{a}_i\|}{d_0} + \delta_i + n_{i,k}
 \label{eq:rss}
\end{equation}
where
\begin{equation}
 \delta_i =
    \begin{cases}
        0, & i\in\mathcal{H}\\
        \delta, & i\in\mathcal{M}
    \end{cases},
    \nonumber
\end{equation}
and $\mathcal{M}$ and $\mathcal{H}$ are, respectively, the set of malicious and honest nodes, $P_0$ is the RSS at a short reference distance $d_0$ (for simplicity referred to as the transmitted power), $\gamma$ is the path loss exponent (PLE) representing the decay of the signal strength with distance, $n_{i,k}$ is the noise term modeled as $n_{i,k}\sim\mathcal{N}(0,\sigma^2_{i,k})$, and $\delta_i \in \mathbb{R}$ is the intensity of the spoofing attack of the $i$-th anchor. Note that in this work the measurement acquisition is assumed practically instantaneous; thus, ambient conditions do not change significantly during the measurement process, having very little to no impact on the deviation of the RSS measurements.

Note that, in contrast to~\cite{Beko:2021}, where malicious anchors could only enlarge their distance measurements, in this work it is assumed that the attackers can either reduce or enlarge distance measurements. For simplicity, it is assumed that the variance of all measurements is equal (for every link and sample), i.e., for $\sigma^2_{i,k}=\sigma^2, i=1, ...,N \text{ and } k=1, ...,K$. Moreover, to easier combat outliers and also for the sake of notation simplicity and without loss of generality, the median of all $K$ RSS measurements in~\eqref{eq:rss} from the $i$-th anchor ($P_i$) is used in the following derivations.

Figure~\ref{fig:uncoord_att} shows an uncoordinated attack, where two (represented by red squares) of the five anchors are malicious and report, independently, false distance measurements, aggravating the localization process. In the concrete case, both malicious anchors reduced their measurements independently from each other, resulting in two intersection points of the red dashed circles to be relatively remote from the main cluster of the intersection points and thus, from the true target location. 

\subsection{Coordinated Attack}
\label{subsec:coord_att}

In coordinated attacks the malicious anchors communicate with each other to agree on a (false) location for the target. The idea is to make the network believe that the target is at a different location than it actually is. Similar to \cite{Mukhopadhyay:2021}, the coordinated attack can be modeled as
\begin{equation}
P_{i,k}=
    \begin{cases}
    \begin{alignedat}{2}
        &P_0-10\gamma\log_{10}\left(\frac{\|\boldsymbol{x}-\boldsymbol{a}_i\|}{d_0}\right)+n_{i,k}\\
        &P_0-10\gamma\log_{10}\left(\frac{\|\boldsymbol{x}_{att}-\boldsymbol{a}_i\|}{d_0}\right)+n_{i,k}
    \end{alignedat}
    \end{cases}
    \hspace{-5.9mm}\text{\quad\parbox{0.4\linewidth}{$i\in\mathcal{H}$\\ \\$i\in\mathcal{M}$,}}
\label{eq:rss_coord}
\end{equation}
where $\boldsymbol{x}_{att}$ is the false location that the attackers agree upon.

\begin{figure}
\begin{center}
\begin{subfigure}{.3\textwidth}
\hspace*{0mm}\includegraphics[width=\textwidth]{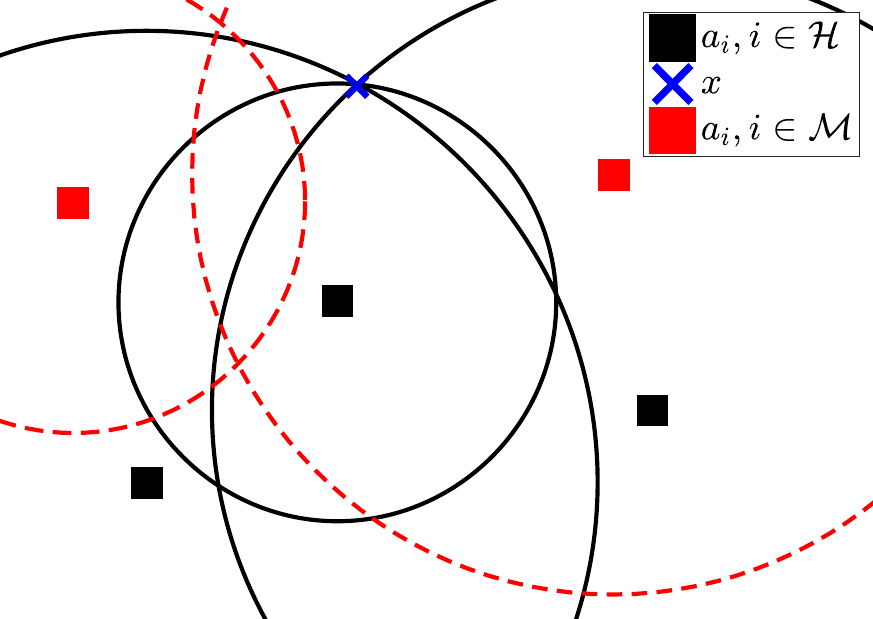}
\caption{Uncoordinated attack}
\label{fig:uncoord_att}
\end{subfigure}
\vspace*{-0mm}
\begin{subfigure}{.3\textwidth}
\hspace*{-0mm}\includegraphics[width=\textwidth]{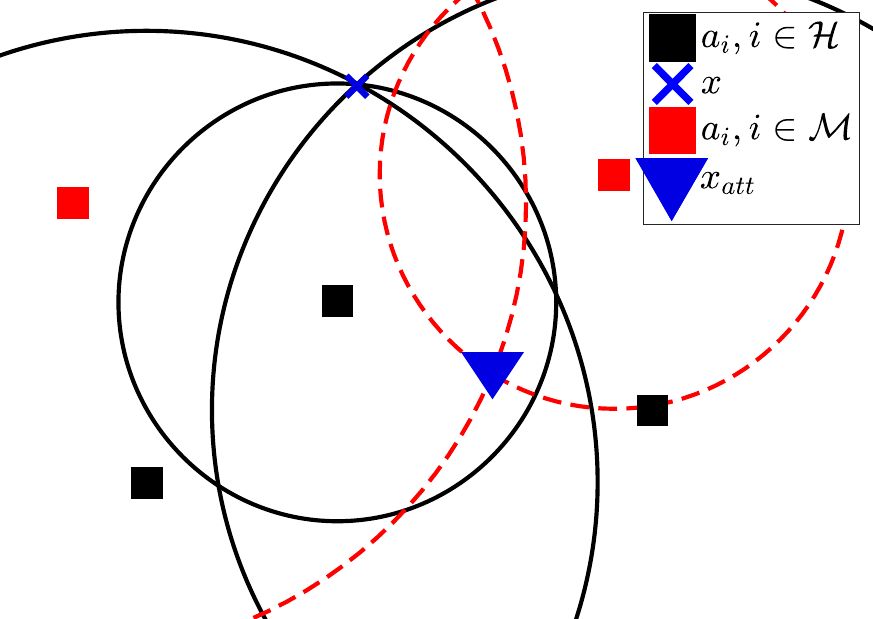}
\caption{Coordinated attack}
\label{fig:coord_att}
\end{subfigure}
\end{center}
\vspace*{-0mm}
\caption{Illustration of different types of spoofing attacks: the red dashed and black solid circles respectively denote malicious and genuine measurements (converted into distance).}
\label{fig:fig:range-based}
\end{figure}

Figure~\ref{fig:coord_att} shows an example of a coordinated attack, where two malicious anchors (represented by a red square) attempt to make the network think that the location of the target is the one represented by a blue triangle ($\boldsymbol{x}_{att}$) instead of the real target position represented by a blue cross ($\boldsymbol{x}$).

Note that~\eqref{eq:rss_coord} is a special case of~\eqref{eq:rss}, having $\delta_i = 10 \gamma \log_{10}\frac{\|\boldsymbol{x}-\boldsymbol{a}_i\|}{\|\boldsymbol{x}_{att}-\boldsymbol{a}_i\|}$. Hence, this work adopts~\eqref{eq:rss} as the general model for both types of spoofing attacks. From it, the probability density function is given by
\begin{equation}
\begin{split}
&f(\boldsymbol{x}; P_i) = \frac{1}{\sqrt{2\pi \sigma^2}} \times\\
&\times\hspace*{-0.5mm}\exp\hspace*{-0.5mm}\left\{\hspace*{-1mm}-\frac{\left( P_{i} - P_0 + 10\gamma\log_{10}{\left(\frac{\|\boldsymbol{x}-\boldsymbol{a}_i\|}{d_0}\right)}-\delta_i \right)^2}{2\sigma^2}\hspace*{-0.75mm}\right\}\hspace*{-0.75mm}
\label{eq:rss_pdf},
\end{split}
\end{equation}
where $\exp\{ \bullet \}$ denotes the exponential function. Minimizing~\eqref{eq:rss_pdf} leads to the ML estimator~\cite{Kay:1993}. However, the ML estimator is non-convex and therefore difficult to tackle directly. In this work, a VS algorithm is introduced to estimate the target's location instead.



\section{The Proposed Approach for the Secure Localization Problem}
\label{sec:proposed_method}

This section describes the derivation of the proposed algorithm for secure localization. It is organized into three parts: 1) a preliminary part, where points of interest are determined, and two main parts in which 2) the proposed localization estimator is described in detail, and 3) the proposed detection procedure to identify attackers is introduced.

\subsection{Determining Points of Interest}
\label{subsec:int_points}

\begin{figure}
\begin{center}
\begin{subfigure}{.325\textwidth}
\hspace*{0mm}\includegraphics[width=\textwidth]{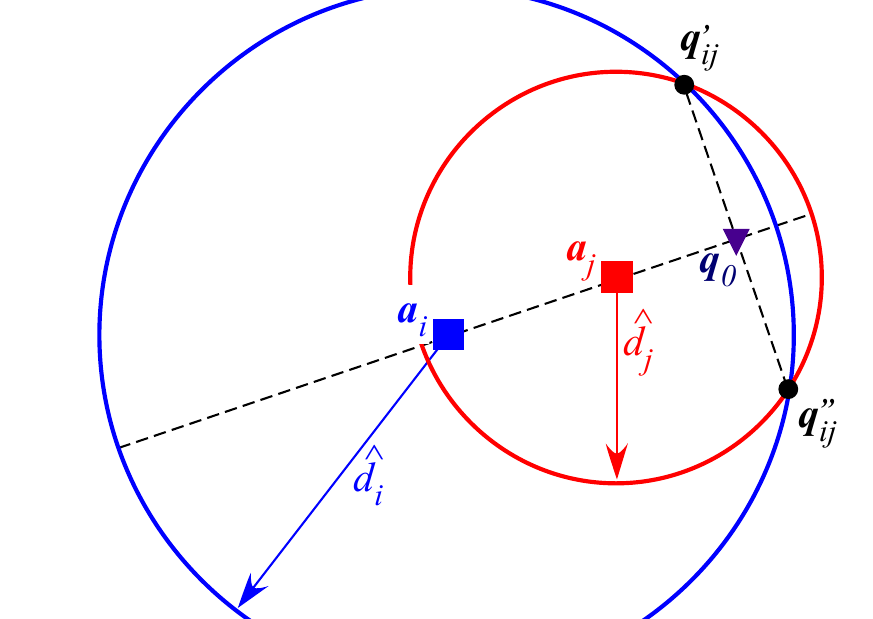}
\caption{True intersections between a pair of anchors}
\label{fig:a}
\end{subfigure}
\vspace*{0mm}
\begin{subfigure}{.325\textwidth}
\hspace*{-0mm}\includegraphics[width=\textwidth]{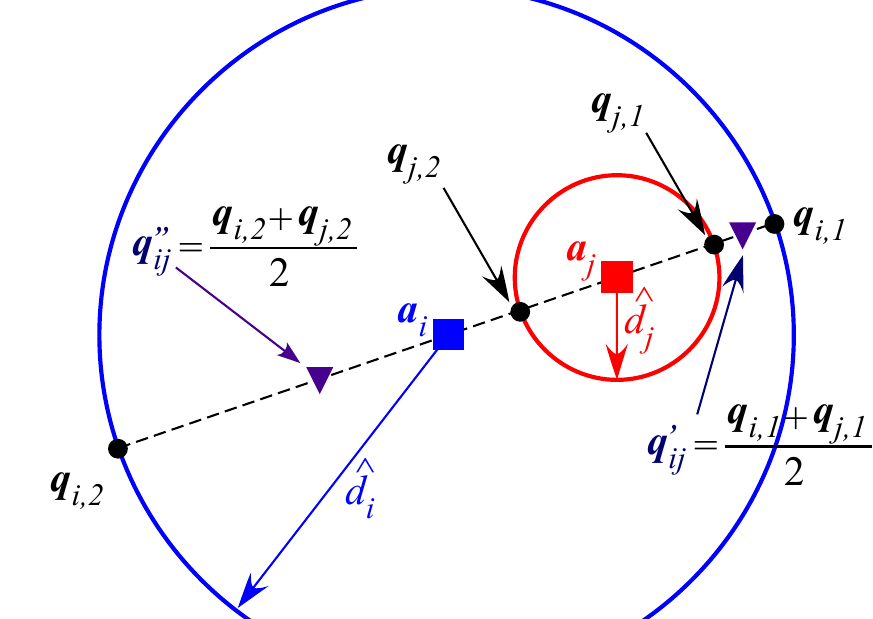}
\caption{Forged intersections between a pair of anchors}
\label{fig:b}
\end{subfigure}
\end{center}
\vspace*{-2mm}
\caption{Illustration of finding/forging circles' intersections.}
\label{fig:intersections}
\end{figure}

At the beginning, all anchors are treated as honest. Hence, the set of malicious nodes is initialized as $\mathcal{M}=\varnothing$, whereas the set of honest nodes is $\mathcal{H}=\{i:1\leq i\leq N\}$. 
Afterwards, one can construct circles, $c_i$, centered at the known locations of the anchors and radii equivalent to the distance estimate, $\widehat{d}_{i}=d_010^{\frac{P_0-P_{i}}{10\gamma}}$, obtained from \eqref{eq:rss}, of the respective anchor (the reader is referred to Fig.~\ref{fig:a}). The intersection points between all pairs of circles are used as points of interest for the voting scheme. The intersection points between a pair of circles (given that they exist) can be calculated~\cite{TomicS:2022} as follows
\begin{equation}
\begin{split}
    \boldsymbol{q}_{ij}'=\boldsymbol{q}_0+\boldsymbol{t}\text{ and } \boldsymbol{q}_{ij}''=\boldsymbol{q}_0-\boldsymbol{t} \text{, for}&\text{ $i=1,...,N-1$,}\\
    &\text{ $j=i+1,...,N$,}
    \label{eq:intersectios}
    \end{split}
\end{equation}
where
\begin{equation}
\nonumber
    \boldsymbol{q}_0=(\boldsymbol{a_j}-\boldsymbol{a_i})\frac{\widehat{d_i}^2-\widehat{d_j}^2}{2\|\boldsymbol{a_j}-\boldsymbol{a_i}\|^2} + \frac{\boldsymbol{a_i}+\boldsymbol{a_j}}{2},
\end{equation}
\begin{equation}
\nonumber
    \boldsymbol{t}=\frac{\sqrt{u}}{2\|\boldsymbol{a_j}-\boldsymbol{a_i}\|^2}\boldsymbol{T}(\boldsymbol{a_j}-\boldsymbol{a_i}) \text{ , } \boldsymbol{T}=
    \begin{bmatrix}
    0 & -1\\
    1 & 0
    \end{bmatrix},
\end{equation}
with
\begin{equation}
\nonumber
u=\left((\widehat{d_i}+\widehat{d_j})^2-\|\boldsymbol{a_j}-\boldsymbol{a_i}\|^2\right)\left(\|\boldsymbol{a_j}-\boldsymbol{a_i}\|^2-(\widehat{d_j}-\widehat{d_i})^2\right),
\end{equation}
according to Fig.~\ref{fig:a}. However, due to the presence of noise and possibly malicious/faulty nodes, a pair of circles might not intersect (the reader is referred to Fig.~\ref{fig:b}). In this case, the tuple set of anchors without intersection is defined as $\mathcal{C}=\{(i,j): c_i\cap c_j = \varnothing\}$, where the notation $c_i\cap c_j=\varnothing$ is used to denote that the circles corresponding to the $i$-th and $j$-th anchors do not intersect. In that case, one can draw a line that passes through the respective pair of anchors, and compute the intersections between the circles and the drawn line as follows.
\begin{equation}
    \begin{split}
    \nonumber
    &\boldsymbol{q}_{i,1}=\boldsymbol{a}_0+\left[(\boldsymbol{a}_i-\boldsymbol{a}_0)^T\boldsymbol{\widehat b} \right. \\
    &\left. +\sqrt{[(\boldsymbol{a}_i-\boldsymbol{a}_0)^T\boldsymbol{\widehat b}]^2-(\boldsymbol{a}_0^T\boldsymbol{a}_0+\boldsymbol{a}_i^T\boldsymbol{a}_i-d_i-2\boldsymbol{a}_0^T\boldsymbol{a}_i)}\right],\\
    &\boldsymbol{q}_{i,2}=\boldsymbol{a}_0+\left[(\boldsymbol{a}_i-\boldsymbol{a}_0)^T\boldsymbol{\widehat b} \right. \\
    &\left. -\sqrt{[(\boldsymbol{a}_i-\boldsymbol{a}_0)^T\boldsymbol{\widehat b}]^2-(\boldsymbol{a}_0^T\boldsymbol{a}_0+\boldsymbol{a}_i^T\boldsymbol{a}_i-d_i-2\boldsymbol{a}_0^T\boldsymbol{a}_i)}\right],\\
    &\boldsymbol{q}_{j,1}=\boldsymbol{a}_0+\left[(\boldsymbol{a}_j-\boldsymbol{a}_0)^T\boldsymbol{\widehat b} \right. \\
    &\left. \sqrt{[(\boldsymbol{a}_j-\boldsymbol{a}_0)^T\boldsymbol{\widehat b}]^2-(\boldsymbol{a}_0^T\boldsymbol{a}_0+\boldsymbol{a}_j^T\boldsymbol{a}_j-d_j-2\boldsymbol{a}_0^T\boldsymbol{a}_j)}\right],\\    
    &\boldsymbol{q}_{j,2}=\boldsymbol{a}_0+\left[(\boldsymbol{a}_j-\boldsymbol{a}_0)^T\boldsymbol{\widehat b} \right. \\
    &\left. -\sqrt{[(\boldsymbol{a}_j-\boldsymbol{a}_0)^T\boldsymbol{\widehat b}]^2-(\boldsymbol{a}_0^T\boldsymbol{a}_0+\boldsymbol{a}_j^T\boldsymbol{a}_j-d_j-2\boldsymbol{a}_0^T\boldsymbol{a}_j)}\right],\\
    \end{split}
\end{equation}
where $\boldsymbol{a}_0=\frac{\boldsymbol{a}_i+\boldsymbol{a}_j}{2}$ is the position vector of the line, and $\boldsymbol{\widehat b}=\frac{(\boldsymbol{a}_i-\boldsymbol{a}_j)}{\|\boldsymbol{a}_i-\boldsymbol{a}_j\|}$ is the unit vector that describes the line's direction. Afterwards, the intersection points to be used in the voting-scheme, $\boldsymbol{q}_{ij}'$, $\boldsymbol{q}_{ij}''$, are obtained as 
\begin{equation}
    \boldsymbol{q}_{ij}'=\frac{\boldsymbol{q}_{i,1}+\boldsymbol{q}_{j,1}}{2}, \,\,\,
    \boldsymbol{q}_{ij}''=\frac{\boldsymbol{q}_{i,2}+\boldsymbol{q}_{j,2}}{2}
    \label{eq:forged_intersections}
\end{equation}

This idea has already been implemented in~\cite{TomicS:2022}, and the reasoning behind it is that when a pair of anchors is genuine, the intersection points would lay in the vicinity of the drawn line, making these forged intersection points a reasonable approximation of the real ones.

\subsection{The Proposed Voting-based Scheme for Target Localization}
\label{subsec:VS}

\begin{figure}
\centering
\includegraphics[width=.35\textwidth]{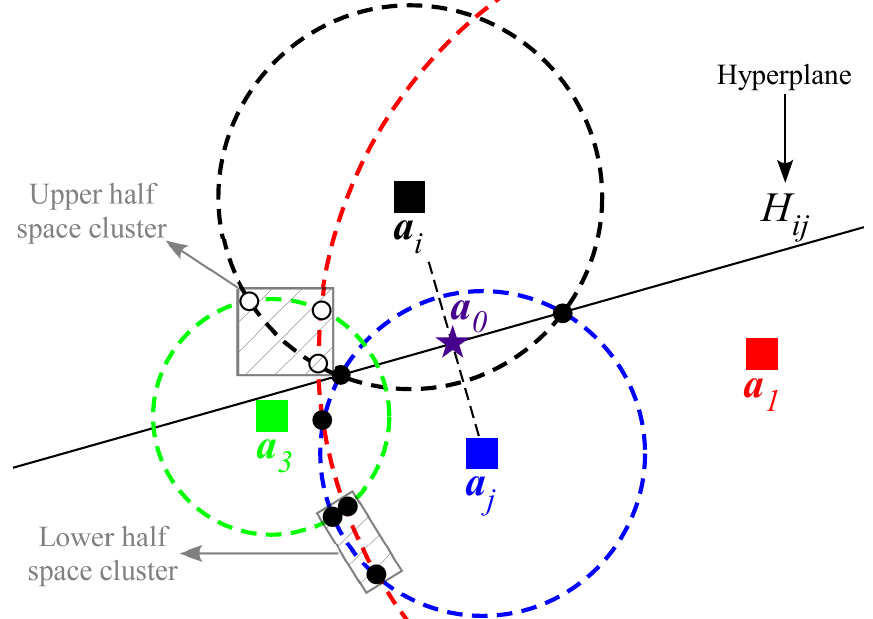}
\caption{The voting process between two anchors when $N = 4$.}
\label{fig:covex_hull}
\vspace{-3mm}
\end{figure}

The voting scheme is a process to cluster and assign votes to the intersection points based on some criterion (for instance, their physical proximity). The main idea is to assign a value (vote) to each intersection point in order to find the most trustworthy ones. It is important to highlight that the localization process is done for a specific instant in time, therefore, past information of the network is not used to aid the voting-scheme. For the sake of simplicity, let us define the matrix $\boldsymbol{Q}=[\boldsymbol{q}_{ij}' \,\, \boldsymbol{q}_{ij}'']\in\mathbb{R}^{2\times\binom{N}{2}}$ containing all intersection points, and the vector $\boldsymbol{Q}_g\in\mathbb{R}^2$ as the $g$-th column of $\boldsymbol{Q}$. This process iterates all pairs of anchors, and for each pair  a hyperplane is computed as 
\begin{equation}
H_{ij}=\{g: \widehat{\boldsymbol{b}}^T \boldsymbol{Q}_g=\widehat{\boldsymbol{b}}^T \boldsymbol{a}_0\},
\label{eq:hyperplane}
\end{equation}
which divides the problem space into two half spaces. The intersection points are assigned to the upper half space, $H_{ij}^{(u)}=\{g: \widehat{\boldsymbol{b}}^T \boldsymbol{Q}_g>\widehat{\boldsymbol{b}}^T \boldsymbol{a}_0\}$, or to the lower half space, $H_{ij}^{(l)}=\{g: \widehat{\boldsymbol{b}}^T \boldsymbol{Q}_g<\widehat{\boldsymbol{b}}^T \boldsymbol{a}_0\}$, according to their physical location with respect to the hyperplane. Next, clusters composed of $N-1$ elements that are physically the closest to each other in each half space are built: $C_{ij}^{(u)}\subseteq H_{ij}^{(u)}$ as the upper cluster and $C_{ij}^{(l)}\subseteq H_{ij}^{(l)}$ as the lower cluster. Lastly, votes, $v_{g}$, are assigned to the points that belong to a cluster (if these exist), based on their distance to the hyperplane, see Figure \ref{fig:covex_hull}. Vote for the $g$-th intersection point is calculated as
\begin{equation}
\begin{split}
    &v_{g} = \displaystyle\sum_{i = 1}^{N-1} \displaystyle\sum_{j = i+1}^{N} w_g\frac{proj_{H_{ij}}(\boldsymbol{Q}_{g})}{\sum_{g:g\in C_{ij}^{(u)}} proj_{H_{ij}}(\boldsymbol{Q}_{g})} +\\
    &+w_g\frac{proj_{H_{ij}}(\boldsymbol{Q}_{g})}{\sum_{g:g\in C_{ij}^{(l)}} proj_{H_{ij}}(\boldsymbol{Q}_{g})}, \, g \in C_{ij}^{(u)}\cup C_{ij}^{(l)}
    \label{eq:vote}
\end{split}
\end{equation}
where
\begin{equation}
     w_g =
      \begin{cases}
      \frac{\widehat{d_j}}{\widehat{d_i}+\widehat{d_j}},&\text{ if $g\in H_{ij}^{(u)}$}\\
      \frac{\widehat{d_i}}{\widehat{d_i}+\widehat{d_j}},&\text{ if $g\in H_{ij}^{(l)}$}\\
    \end{cases},
\label{eq:halfspace_weights}
\end{equation}
\begin{equation}
\label{eq:projections}
proj_{H_{ij}}(\boldsymbol{Q}_g)=\|\boldsymbol{Q}_g-(\boldsymbol{e}\boldsymbol{e}^T\boldsymbol{Q_g}+(\boldsymbol{I}_2-\boldsymbol{e}\boldsymbol{e}^T)\boldsymbol{a}_0)\|
\end{equation}
with $\boldsymbol{e}=\boldsymbol{T}\boldsymbol{\widehat{b}}$,
$\boldsymbol{I}_2$ is the identity matrix of order two, $proj_{H_{ij}}(\boldsymbol{Q}_g)$ denotes the distance of an intersection point, $\boldsymbol{Q}_g$, to it's respective projection on the hyperplane, and $w_g$ is a weight based on the distances $\widehat{d_i}$ and $\widehat{d_j}$ which takes into account the anchor and the point $\boldsymbol{Q}_g$.

The procedure to compute the votes defined in~\eqref{eq:vote} is explained in the following. For every point of interest $\boldsymbol{Q}_g$, with $g = 1, ..., \binom{N}{2}$, one considers all combinations of pairs of anchors $(i,j)$, where $i = 1, ..., N$ and $j = i + 1, ..., N$, and calculates the respective hyperplanes $H_{ij}$ according to~\eqref{eq:hyperplane}. Then, each weight $w_g$ divides a single vote according to the proximity of $H_{ij}$ to $\boldsymbol{a}_i$ and $\boldsymbol{a}_j$, such that more weight is assigned to the half space containing the anchor closer to the hyperplane~\eqref{eq:halfspace_weights}. A cluster of $N-1$ points physically closest to one another, $C_{ij}^{(u)}$ and $C_{ij}^{(l)}$, are formed in each half space. If $g \in C_{ij}^{(u)}$ or $g \in C_{ij}^{(l)}$, distances of the cluster points to $H_{ij}$ are calculated~\eqref{eq:projections} and converted into weights by dividing the individual distances with the sum of all distances of the points in the cluster to the hyperplane; otherwise, sum zero and move on to the next anchor pair. Lastly, these weights are further weighted with $w_g$ and summed up to form a vote for the $g$-th point of interest. Hence, the intuition behind the votes defined in~\eqref{eq:vote} is that it assigns greater weights (belief) to points closer to the hyperplane, since the correct cluster of (genuine) points should lie in its vicinity.
    
An estimate of the target's location can be obtained by re-ordering the vote vector in a descending fashion, $\Tilde{\boldsymbol{v}}=[\Tilde{v}_h]$, such that $\Tilde{{v}}_1\geq\Tilde{{v}}_2\geq\ldots\geq\Tilde{{v}}_{\binom{N}{2}}$. The first $N-1$ votes correspond to the most trustworthy points; hence, the estimate is obtained by applying the WCM principle for the $N-1$ (normalized) vote values as
\begin{equation}
    \widehat{\boldsymbol{x}}=\sum_{h=1}^{N-1} \Tilde{w}_{h}\boldsymbol{Q}_{h}, \,\,\, \text{with} \,\,\, \Tilde{w}_{h}=\frac{\Tilde{v}_h}{\sum_{h=1}^{N-1} \Tilde{v}_h}.
    \label{eq:estimate}
\end{equation}

\subsection{Attack Detection}
\label{subsec:detection}
Considering \eqref{eq:rss_pdf}, in the case of a malicious anchor, it is intuitive that the attack would shift the probability distribution according to the attack intensity. This shift of the distribution could cause the mean to fall outside of a confidence interval. In other words, one could take advantage of it to detect attackers. However, as the attack intensity is unknown, it can be estimated by exploiting the estimated target location in~\eqref{eq:estimate} based on the ML principle. Similar can be done for the noise standard deviation and the expected (genuine) RSS value as
\begin{equation}
    \widehat{\delta_i}=\frac{\sum_{k=1}^{K} P_0 - P_{i,k}-10\gamma\log_{10}{\|\boldsymbol{\widehat{\boldsymbol{x}}-\boldsymbol{a}_i}\|}}{K},
    \label{eq:attack_estimate}
\end{equation}
\begin{equation}
    \widehat{\sigma}=\frac{1}{N}\sum_{i=1}^N\sqrt{\frac{1}{(K-1)}\sum_{k=1}^{K} \left(\alpha_i-\widehat{\delta_i}\right)^2}\text{,}
	\label{eq:std_est}
\end{equation}

\begin{equation}
    \widehat{P}_i = P_0 - 10\gamma\log_{10}{\|\boldsymbol{\widehat{\boldsymbol{x}}-\boldsymbol{a}_i}\|},
    \label{eq:pi_estimate}
\end{equation}
with $\alpha_{i,k} = P_{i,k} - P_0 + 10\gamma\log_{10}{\|\boldsymbol{\widehat{\boldsymbol{x}}-\boldsymbol{a}_i}\|}$.

It is well known that 68~\% of the mass of a normal distribution falls within one standard deviation of the mean. Thus, if the measured RSS in~\eqref{eq:rss} lays outside of the confidence interval $[\widehat{P}_i-\widehat{\sigma}, \widehat{P}_i+\widehat{\sigma}]$, the respective anchor is classified as malicious, and the set of malicious nodes becomes $\mathcal{M}=\{i: P_i < \widehat{P}_i - \widehat{\sigma} \lor P_i > \widehat{P}_i + \widehat{\sigma}\}$.

Note that, in contrast to \cite{Beko:2021}-\cite{Mukhopadhyay:2021}, the malicious node is possibly exploited in the localization process. This can be advantageous when the $|\delta_i|$ is not high compared to noise power, and quantity of anchors prevails over quality.

The proposed algorithm is summarized in Algorithm~\ref{al:VS}.
\begin{algorithm}\footnotesize
\scriptsize
\caption{~Pseudo-code for the Proposed VS Algorithm}
\begin{algorithmic}[1]
\REQUIRE $N:$ Number of anchors in the network
\REQUIRE $\boldsymbol{a}_i:$ True anchor locations $i = 1, ..., N$
\REQUIRE $K:$ Number of measurement samples
\REQUIRE $\gamma:$ Path loss exponent
\REQUIRE $d_0:$ Reference distance
\REQUIRE $P_0:$ Transmit power
\REQUIRE $P_{i,k}:$ $k$-th RSS measurement sample at $i$-th anchor
\STATE \textbf{Initialization:} Set $\mathcal{M}=\varnothing$ and $\mathcal{H}=\left\{ i: 1 \leq i \leq N \right\}$
\STATE Form circles centered at $\boldsymbol{a}_i$ with radius $d_i$ for $i = 1, ..., N$
\FOR{$i = 1, ..., N-1$}
	\FOR{$j = i+1, ..., N$}
		\IF{$c_i \cap c_j \neq \varnothing$}
			\STATE $q_{ij}', q_{ij}'' \leftarrow$~\eqref{eq:intersectios} {\fontfamily{qcr}\selectfont //Find circle intersection points}
		\ELSE
			\STATE $q_{ij}', q_{ij}'' \leftarrow$~\eqref{eq:forged_intersections} {\fontfamily{qcr}\selectfont //Forge intersection points}
		\ENDIF
		\STATE $H_{ij} \leftarrow$~\eqref{eq:hyperplane} {\fontfamily{qcr}\selectfont //Compute hyperplanes}
		\STATE $C_{ij}^{(l)} \subseteq H_{ij}^{(l)}$ and $C_{ij}^{(u)} \subseteq H_{ij}^{(u)}$ {\fontfamily{qcr}\selectfont //Cluster in half spaces}
	\ENDFOR
\ENDFOR
\STATE $v_g \leftarrow$~\eqref{eq:vote} {\fontfamily{qcr}\selectfont //Compute votes for cluster points}
\STATE $\widehat{\boldsymbol{x}} \leftarrow$~\eqref{eq:estimate} {\fontfamily{qcr}\selectfont //Estimate target's location}
\STATE $\widehat{\delta}_i \leftarrow$~\eqref{eq:attack_estimate} {\fontfamily{qcr}\selectfont //Estimate attack intensity}
\STATE $\widehat{\sigma} \leftarrow$~\eqref{eq:std_est} {\fontfamily{qcr}\selectfont //Estimate noise STD}
\STATE $\widehat{p}_i \leftarrow$~\eqref{eq:pi_estimate} {\fontfamily{qcr}\selectfont //Estimate expected genuine RSS value}
\FOR{$i = 1, ..., N$}
	\IF{$P_i \in \left[ \widehat{P}_i - \widehat{\sigma}, \, \widehat{P}_i + \widehat{\sigma} \right]$}
		\STATE $\mathcal{M} \leftarrow \mathcal{M} \cup \left\{i\right\}$ {\fontfamily{qcr}\selectfont //Label anchor $i$ as malicious}
		\STATE $\mathcal{H} \leftarrow \mathcal{H} \setminus \left\{i\right\}$ {\fontfamily{qcr}\selectfont //Remove anchor $i$ from \textit{honest} ones}
	\ENDIF
\ENDFOR
\STATE \textbf{Return:} $\widehat{\boldsymbol{x}}$ and $\mathcal{M}$
\end{algorithmic}
\label{al:VS}
\end{algorithm}



\section{Performance Analysis}
\label{sec:performance}

This section presents a series of numerical results in order to assess the performance of the proposed solution. It presents analysis based on computational complexity, localization accuracy and success in detecting malicious attackers, as well as a discussion on the limitations of the proposed scheme. It is thus organized correspondingly.

\subsection{Complexity Analysis}
\label{complexity}

The complexity analysis is highly relevant for the applicability of the algorithm, especially in real-time scenarios. Given that $B_{\text{max}}$ and $B_{\text{ADMM}}$ are respectively the maximum number of iterations for the GTRS-based and for the ADMM-based algorithms, Table~\ref{table:table1} summarizes the worst-case computational complexity together with the average running time of the considered methods. The latter evaluation was performed with $100$ Monte Carlo (MC) runs, for $N = 6$, $\sigma = 1$~dB, $\delta = 10$~dB, on a machine with the following characteristics: CPU: Intel(R) Core(TM) i5-1135G7 CPU @ 2.40 GHz, RAM: 16 GB, OS: Windows 11 Home, running MATLAB R2016b.

Essentially, all considered methods involve matrix addition, multiplication and transpose operations, and the LC-GTRS in~\cite{TomicS:2022} and SWLS in~\cite{Mukhopadhyay:2021} require matrix inversion. These operations come with certain computational costs associated with them (e.g., $\mathcal{O}(m^3)$ is the cost of matrix inversion, where $m$ stands for the size of the square matrix). Nevertheless, one can note from the table that all considered algorithms have linear complexity in the dominant term, $N$. Regarding the complexity of the proposed VS method, its most expensive operation is the calculation of the votes in~\eqref{eq:vote}, where a series of matrix additions and multiplications is required in order to compute the projection of points of interest onto the hyperplane. However, dimensions of the matrices and vectors involved in these computations are fixed to $y \times y$ and $y$ respectively, with $y$ denoting the dimension of the space of interest (2-D or 3-D). Hence, although the operations in VS are computationally the least demanding, the proposed algorithm requires repeated actions (for each pair of anchors), which results in somewhat higher average running time than SWLS and R-GTRS. However, note that these actions could be done in parallel, so that the running time of VS is reduced significantly. Moreover, the GTRS-based and ADMM-based approaches require repeated actions, while LC-GTRS and LN-1E require two phases to obtain the final location estimation.

\begin{table*}\footnotesize
\caption{Brief Summary of the Considered Algorithms}
\vspace*{-4mm}
  \begin{center}
  \begin{adjustbox}{width=0.85\textwidth}
	\small
	\begin{tabular}{|c|c|c|c|c|}
	\hline
	\textbf{Algorithm} & \textbf{Complexity} & \textbf{Running Time~(ms)} & \textbf{Type of attack} & \textbf{Additional requirements}\\ \hline \hline
	VS in Section~\ref{sec:proposed_method} & $\binom{N}{2} \times \mathcal{O}(N)$ & $7$ & UC \& C & No \\ \hline
	LC-GTRS in~\cite{TomicS:2022}  & $2 \times \mathcal{O}(B_{\text{max}} \times N)$  & $5$ & UC \& C & No \\ \hline
    SWLS in~\cite{Mukhopadhyay:2021}  & $\mathcal{O}(N)$ &     $1$  & UC & Knowledge about $\sigma$ \\ \hline
    LN-1E in~\cite{Mukhopadhyay:2021}  & $2 \times \mathcal{O}(B_{\text{ADMM}} \times N)$ & $18$ & C & No \\ \hline
    R-GTRS in~\cite{Tomic_TVT1:2024}  & $\mathcal{O}(B_{\text{{max}}} \times N)$  & $1$ & UC \& C & Knowledge about $\max\{|\delta_i|\}$ \\ \hline
    WADMM in~\cite{Tomic_TVT2:2024}  & $\mathcal{O}(B_{\text{ADMM}} \times N)$  & $20$ & UC \& C & No \\ \hline
	\end{tabular}
   \end{adjustbox}
   \end{center}
\label{table:table1}
\end{table*}

\subsection{Performance Analysis via Simulations}
\label{results}

This section validates the performance of the considered algorithms in terms of localization accuracy and success of attacker detection through numerical simulations. All simulations disclose the results for $N$ randomly deployed anchors and a single target (at a time), within a two-dimensional area of $25\times 25~\text{m}^2$ with two malicious anchors. Moreover, anchors are randomly deployed $N_D=1000$ times and, for each localization setting, two randomly chosen anchors are considered malicious $N_A=50$ times.

The RSS measurements\footnote{Note that this work considers PLE and the transmit power known a priori. This might not be the case in practice, so these parameters might not be (perfectly) known beforehand. Considering these parameters not known might be an interesting direction for future work.} were obtained through \eqref{eq:rss}, where the transmit power at the target node is set to $P_0=15$~dBm, the PLE is set at $\gamma=3$, which corresponds to propagation in an urban area, and $K=10$ observation were considered. The main metric used to assess the localization performance is the root mean squared error (RMSE), $\text{RMSE} = \sqrt{\sum_{i=1}^{Mc}\frac{\| \boldsymbol{x}_i - \boldsymbol{\widehat{x}}_i\|^2}{MC}}$~(m), where $x_i$ and $\widehat{x}_i$ are, respectively, the true and the estimated target location in the $i$-th Monte Carlo run, i.e., $MC = N_D \times N_A$.

It is worth mentioning here that SWLS in~\cite{Mukhopadhyay:2021} requires tuning the detection threshold by studying an empirical parameter, $\zeta$. By fine-tuning the empirical threshold in the considered settings, it was concluded that the best localization results for SWLS were obtained for $\zeta=1.5$; thus, this value is adopted for SWLS in all presented simulations. Furthermore, R-GTRS in~\cite{Tomic_TVT1:2024} requires knowledge on the magnitude of the attack intensity, i.e., $|\delta_i|$ in~\eqref{eq:rss}. The true value $\delta_i$ is given to R-GTRS in all presented simulations.

\subsubsection{Uncoordinated Attacks}
\label{subsucsec:res_uncoord}

In a non-coordinated attack, the attack intensity in~\eqref{eq:rss} is chosen according to
\begin{equation}
\delta_i =
\begin{cases}
0, & i \in \mathcal{H}\\
\delta, & i \in \mathcal{M}
\end{cases},
\nonumber
\end{equation}
where $\delta$~(dB) is defined below for each scenario.

\begin{figure}
\begin{center}
\begin{subfigure}{.425\textwidth}
\hspace*{0mm}\includegraphics[width=\textwidth]{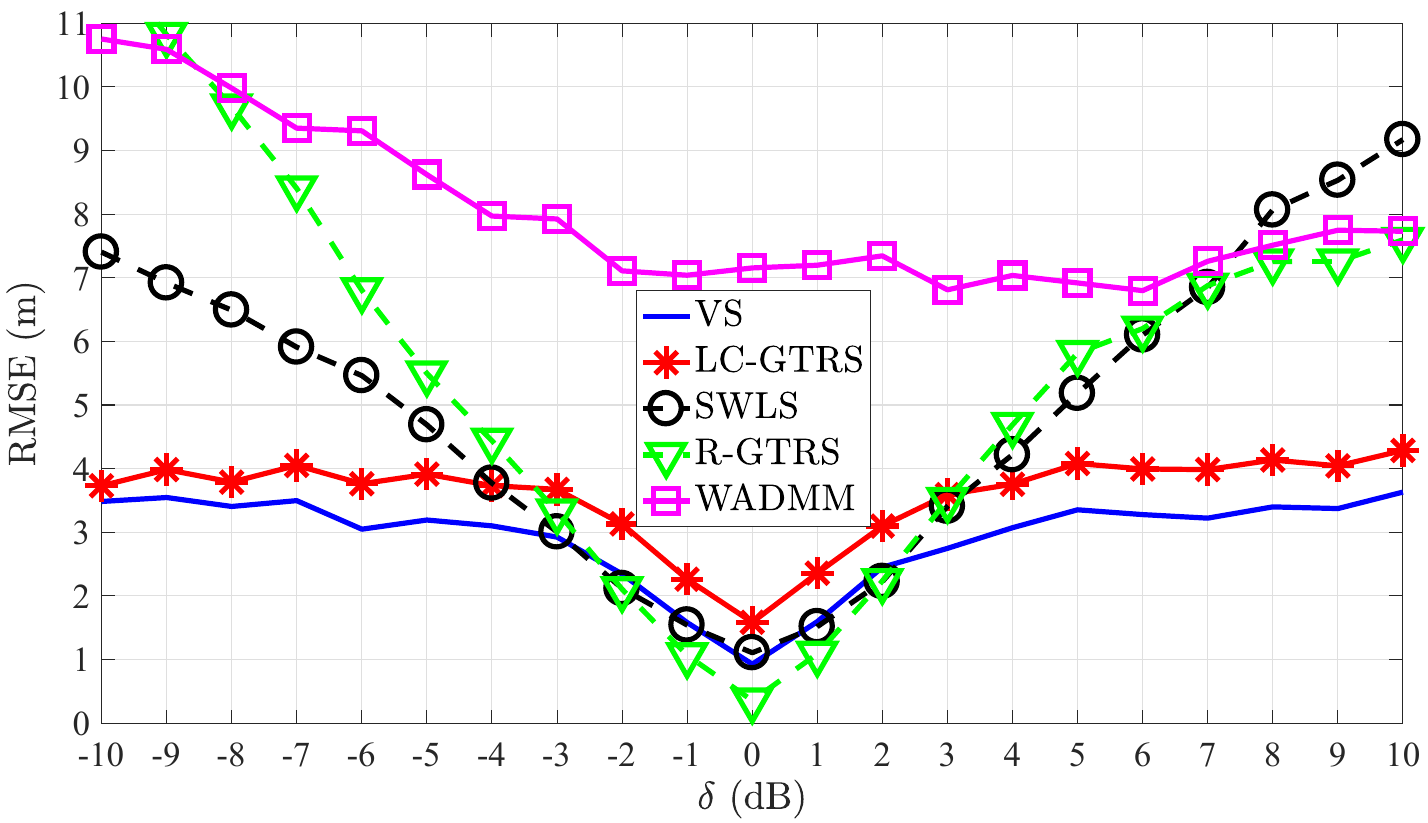}
\caption{$N = 6$ and $\sigma = 1$~(dB)}
\label{fig:RMSE_vs_delta_uncoord_N6_sigma1}
\end{subfigure}
\vspace*{0mm}
\begin{subfigure}{.425\textwidth}
\hspace*{-3mm}\includegraphics[width=\textwidth]{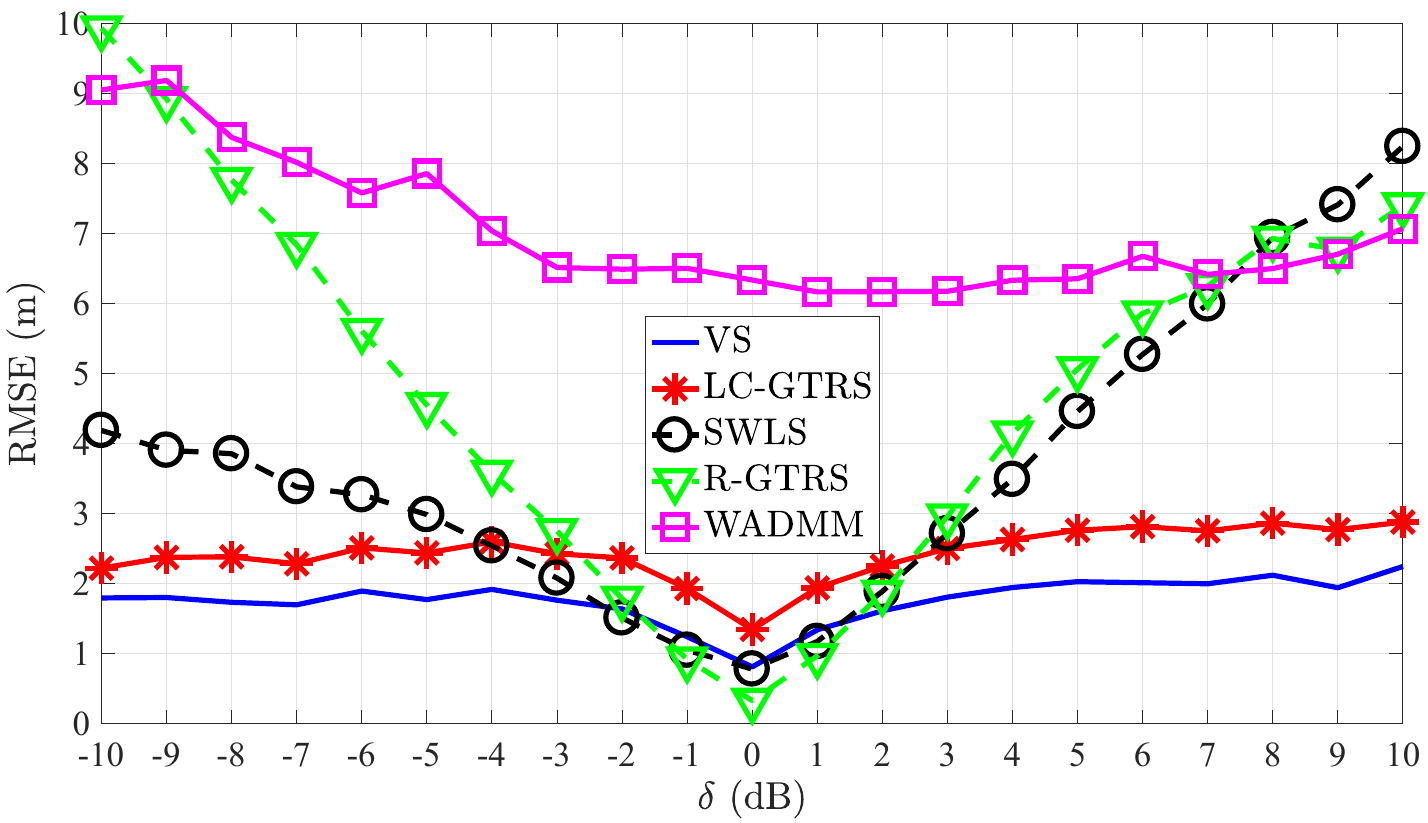}
\caption{$N = 7$ and $\sigma = 1$~(dB)}
\label{fig:RMSE_vs_delta_uncoord_N7_sigma1}
\end{subfigure}
\begin{subfigure}{.425\textwidth}
\hspace*{-3mm}\includegraphics[width=\textwidth]{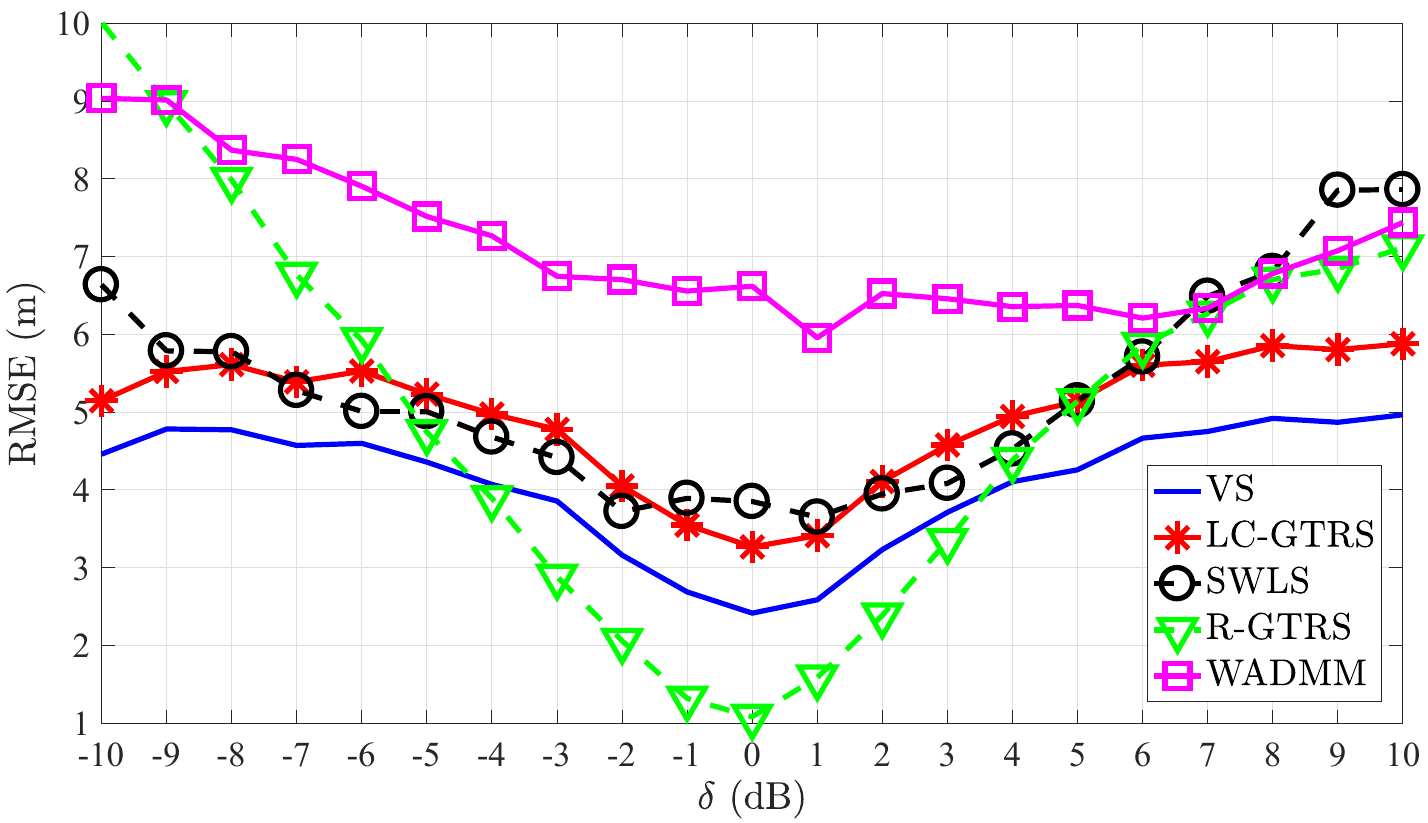}
\caption{$N = 7$ and $\sigma = 3$~(dB)}
\label{fig:RMSE_vs_delta_uncoord_N7_sigma3}
\end{subfigure}
\end{center}
\vspace*{0mm}
\caption{RMSE~(m) versus $\delta$~(dB) in an uncoordinated attack scenario.}
\label{fig:RMSE_vs_delta_uncoord}
\end{figure}

Fig.~\ref{fig:RMSE_vs_delta_uncoord} illustrates RMSE~(m) versus $\delta$~(dB) comparison of the considered approaches in an uncoordinated attack scenario with two malicious anchors for different values of $N$ and $\sigma$~(dB). As expected, the figure clearly shows that the performance of all methods degrades with the increase of $\abs{\delta}$~(dB). Likewise, it shows a trend that all methods benefit from the increase of $N$ and suffer from the increase of $\sigma$~(dB). From Fig.~\ref{fig:RMSE_vs_delta_uncoord}, one can see that the proposed solution outperforms the existing ones for medium-to-high $\abs{\delta}$~(dB), whereas R-GTRS exhibits superior performance for low $\abs{\delta}$~(dB). Nevertheless, R-GTRS suffers significant performance degradation for high $\abs{\delta}$~(dB), making it unusable in these settings.

\begin{figure}
\begin{center}
\begin{subfigure}{.425\textwidth}
\hspace*{0mm}\includegraphics[width=\textwidth]{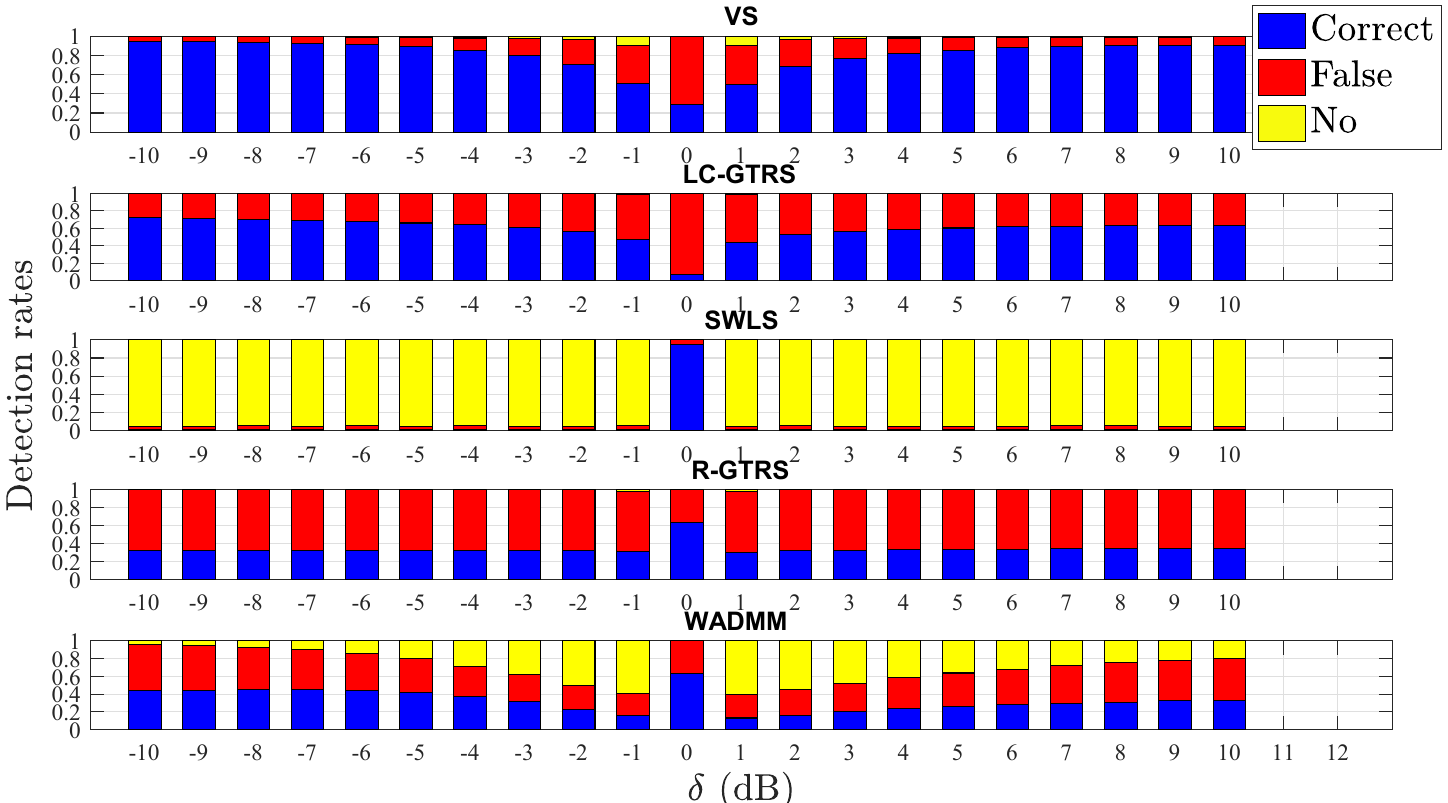}
\caption{$N = 6$ and $\sigma = 1$~(dB)}
\label{fig:DR_vs_delta_uncoord_N6_sigma1}
\end{subfigure}
\vspace*{0mm}
\begin{subfigure}{.425\textwidth}
\hspace*{-3mm}\includegraphics[width=\textwidth]{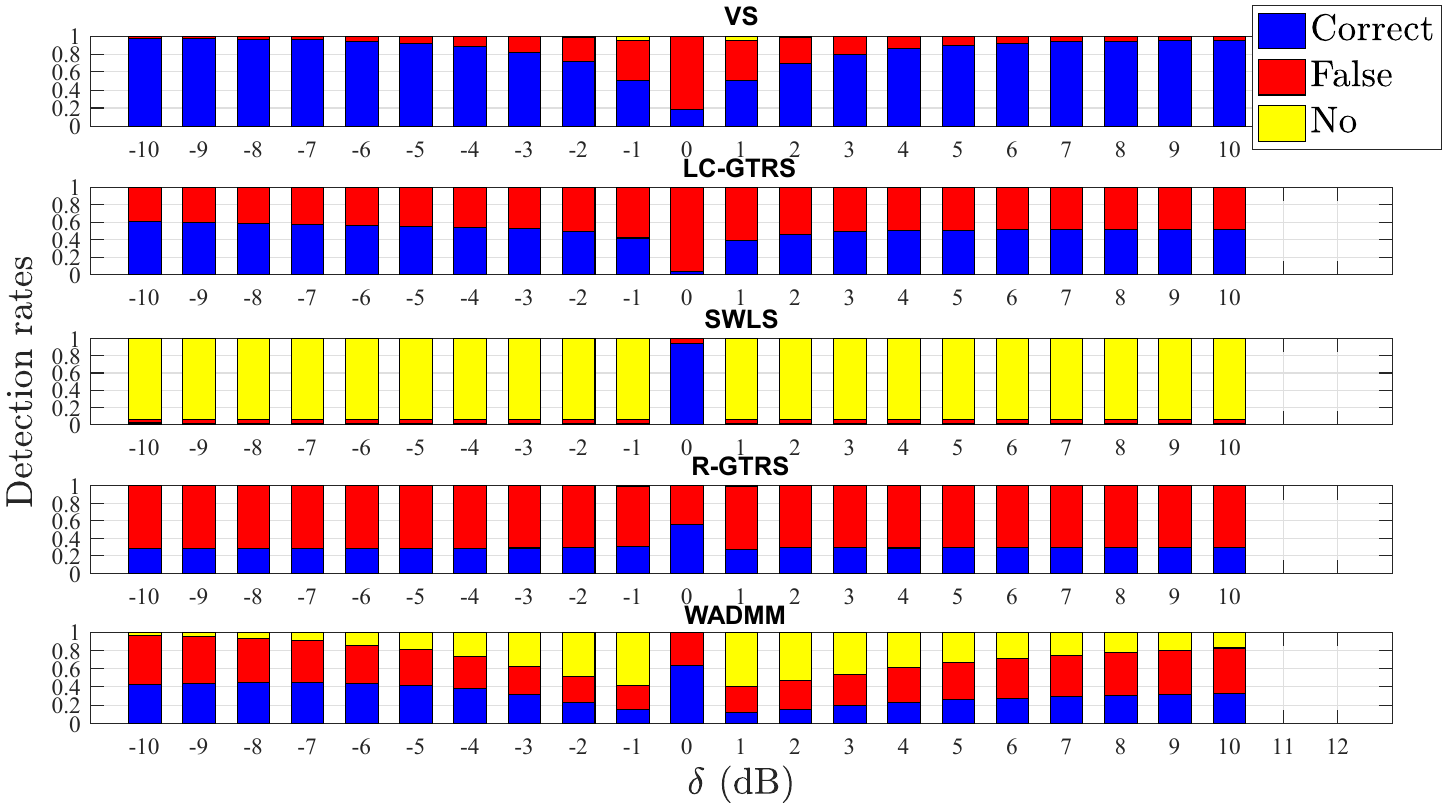}
\caption{$N = 7$ and $\sigma = 1$~(dB)}
\label{fig:DR_vs_delta_uncoord_N7_sigma1}
\end{subfigure}
\begin{subfigure}{.425\textwidth}
\hspace*{-3mm}\includegraphics[width=\textwidth]{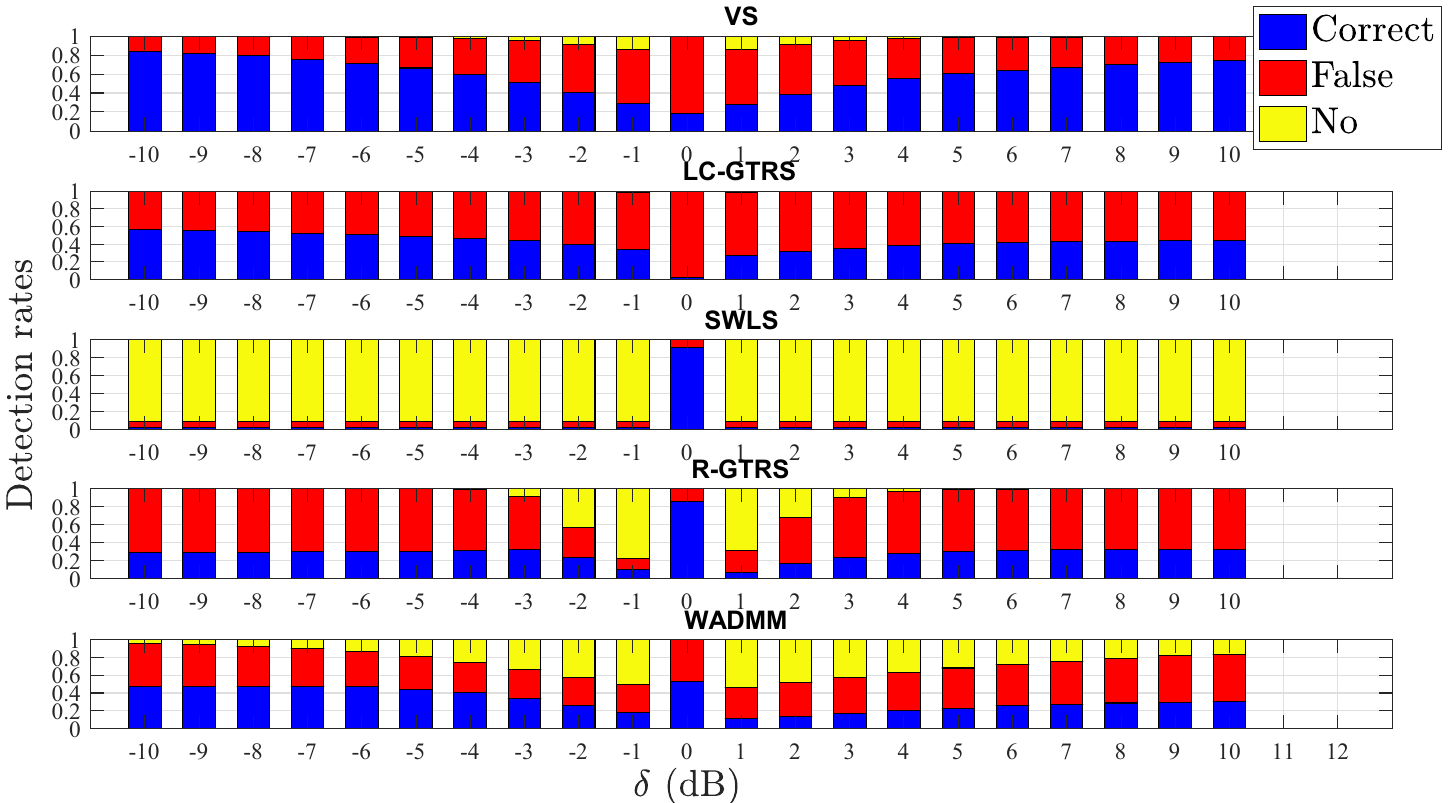}
\caption{$N = 7$ and $\sigma = 3$~(dB)}
\label{fig:DR_vs_delta_uncoord_N7_sigma3}
\end{subfigure}
\end{center}
\vspace*{0mm}
\caption{Detection rates versus $\delta$~(dB) in an uncoordinated attack scenario.}
\label{fig:DR_vs_delta_uncoord}
\end{figure}

Fig.~\ref{fig:DR_vs_delta_uncoord} illustrates different detection rates versus $\delta$~(dB) comparison of the considered approaches in an uncoordinated attack scenario with two malicious anchors for different values of $N$ and $\sigma$~(dB). The figure corroborates the effectiveness of the proposed detection scheme and confirms the superiority of the proposed approach for high $\abs{\delta}$~(dB). The latter claim can be explained by the robustness of the proposed approach to high $\abs{\delta}$~(dB) in terms of localization performance, since the location estimate of VS is exploited for detection, and having better localization estimate can naturally lead better detection performance (if used in a correct manner). Finally, it is worth mentioning that the detection of SWLS is poor in general, which suggest that its hyper-parameter $\zeta$ might need more fine-tuning (even though $\zeta = 1.5$ resulted in the best localization performance).

\begin{figure}
\centering
\includegraphics[width=.425\textwidth]{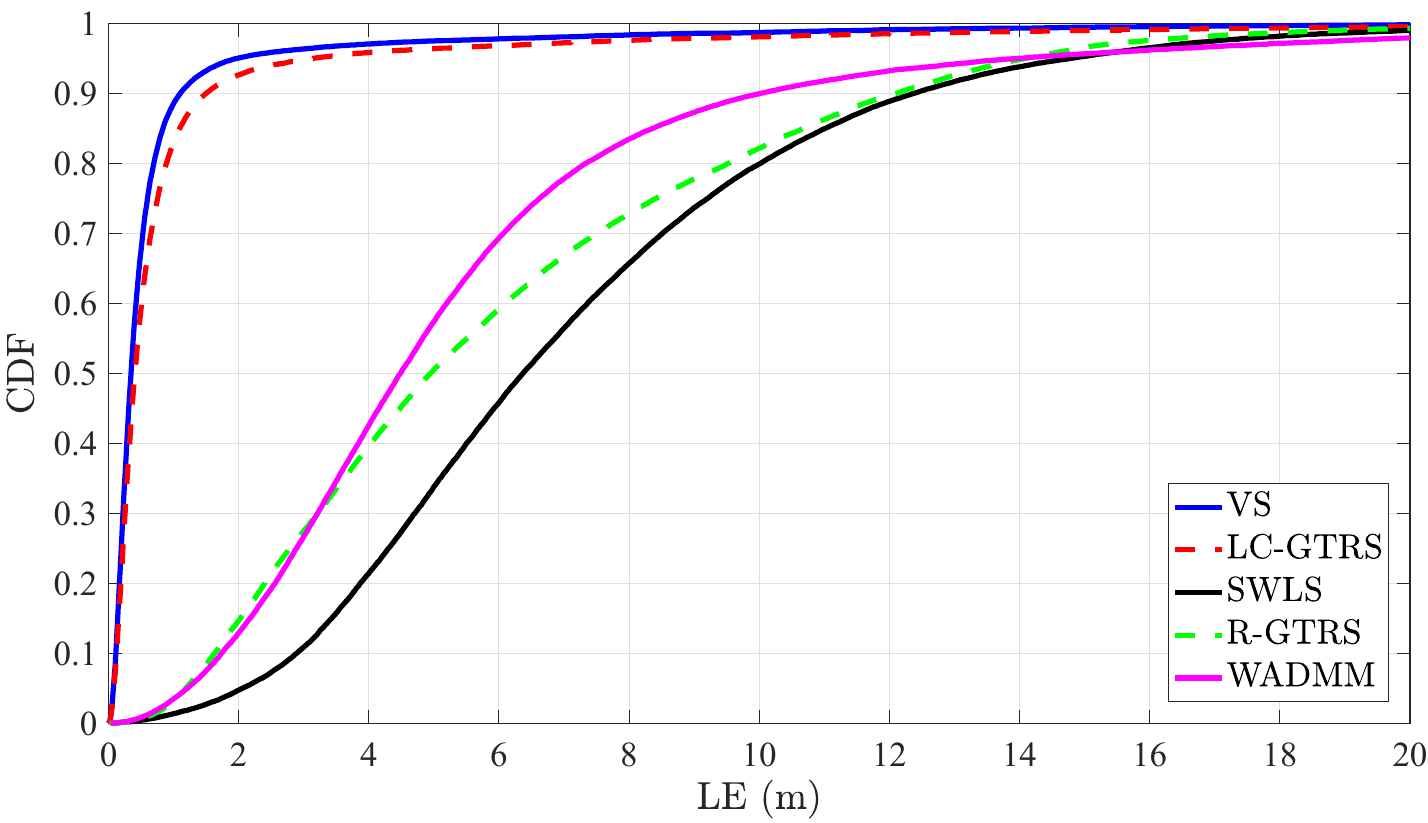}
\caption{CDF versus LE~(m) in an uncoordinated attack scenario, when $N = 7$ and $\sigma = 1$~(dB).}
\label{fig:CDF_vs_LE_uncoord}
\vspace{-3mm}
\end{figure}

Fig.~\ref{fig:CDF_vs_LE_uncoord} illustrates the cumulative distribution function (CDF) versus localization error (LE) in the considered uncoordinated scenario, when $N = 7$, $\delta = 7$~(dB) and $\sigma = 1$~(dB). The LE in the $i$-th MC run is defined as $\text{LE}_i = \| \boldsymbol{x}_i - \boldsymbol{\widehat{x}}_i \|$. The figure shows that the proposed scheme and LC-GTRS exhibit similar performance which is clearly superior over the remaining existing ones. For instance, the former two solutions have a median of $\text{LE} \approx 0.4$~(m), whereas the best of the remaining ones exhibits a median of $\text{LE} \approx 4.5$~(m).

\subsubsection{Coordinated Attacks}
\label{subsubsec:res_coord}

In a coordinated attack, $\boldsymbol{x}_{att}$ in~\eqref{eq:rss_coord} is obtained by choosing a random point on a circle centered at the true target's location with radius $\delta$, i.e., $\boldsymbol{x}_{att} = \boldsymbol{x} + \delta [\cos\theta, \sin\theta]^T$, where $\theta$ is a random angle chosen from a uniform distribution from the interval $[0, 2	\pi]$.

\begin{figure}
\begin{center}
\begin{subfigure}{.425\textwidth}
\hspace*{0mm}\includegraphics[width=\textwidth]{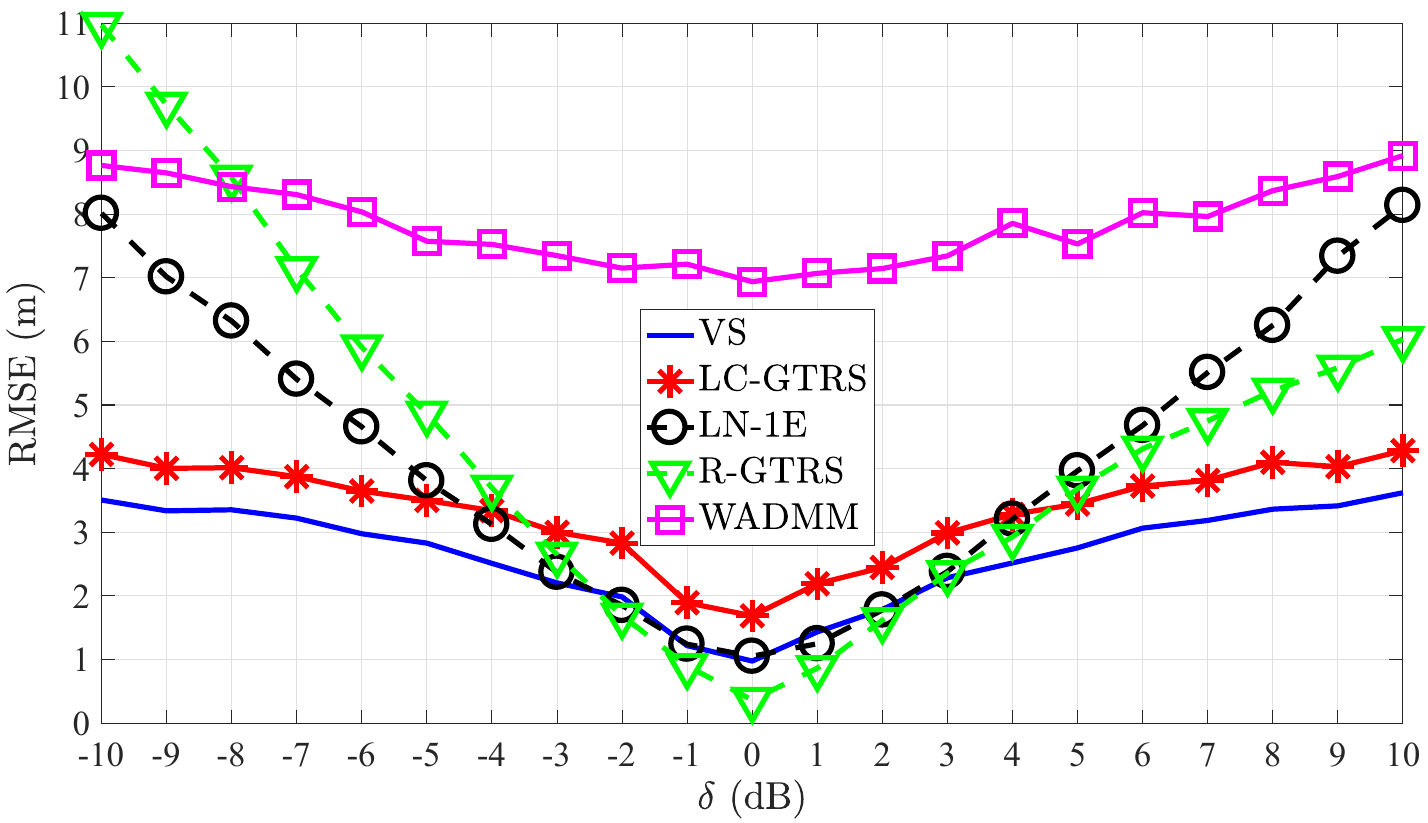}
\caption{$N = 6$ and $\sigma = 1$~(dB)}
\label{fig:RMSE_vs_delta_coord_N6_sigma1}
\end{subfigure}
\vspace*{0mm}
\begin{subfigure}{.425\textwidth}
\hspace*{-3mm}\includegraphics[width=\textwidth]{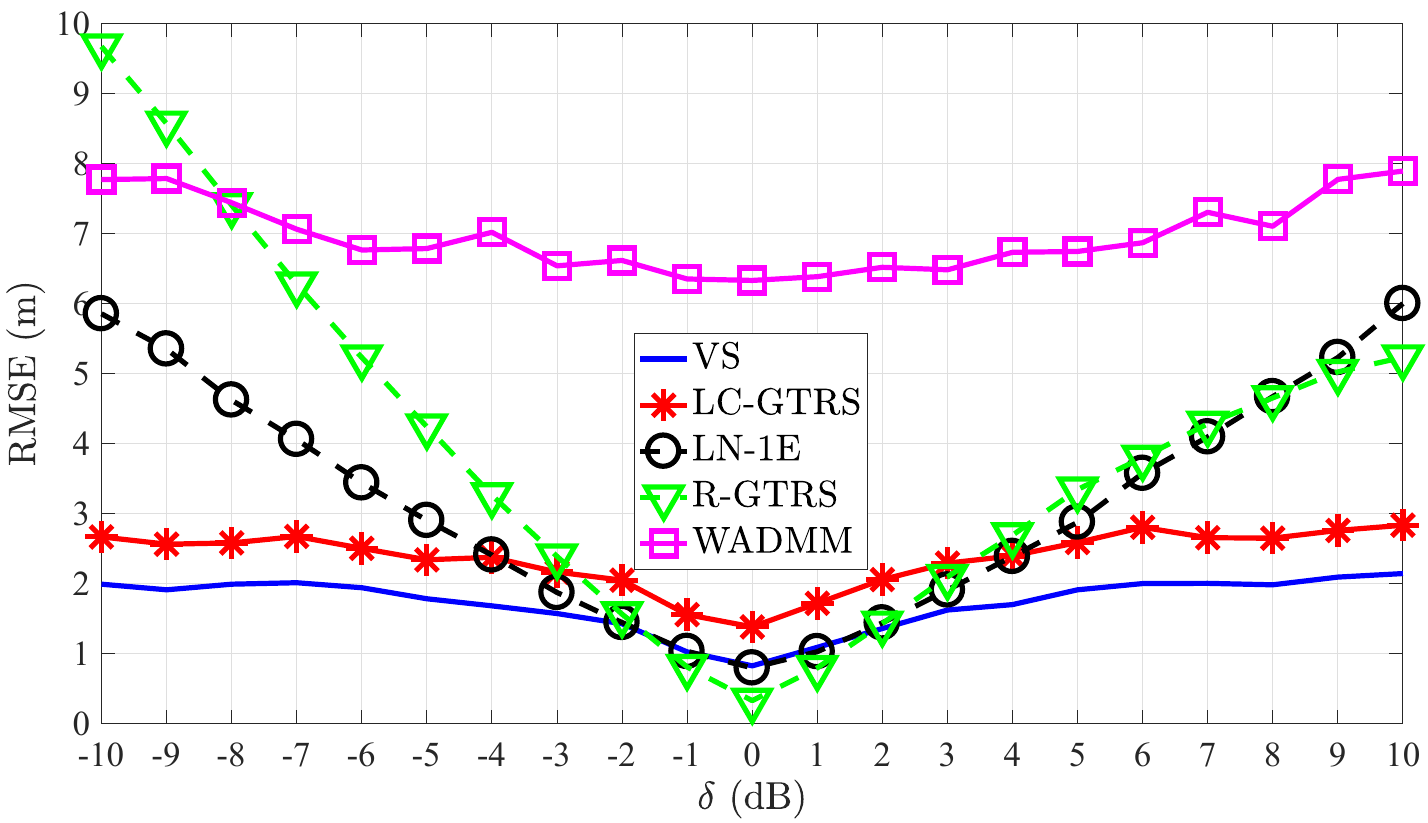}
\caption{$N = 7$ and $\sigma = 1$~(dB)}
\label{fig:RMSE_vs_delta_coord_N7_sigma1}
\end{subfigure}
\begin{subfigure}{.425\textwidth}
\hspace*{-3mm}\includegraphics[width=\textwidth]{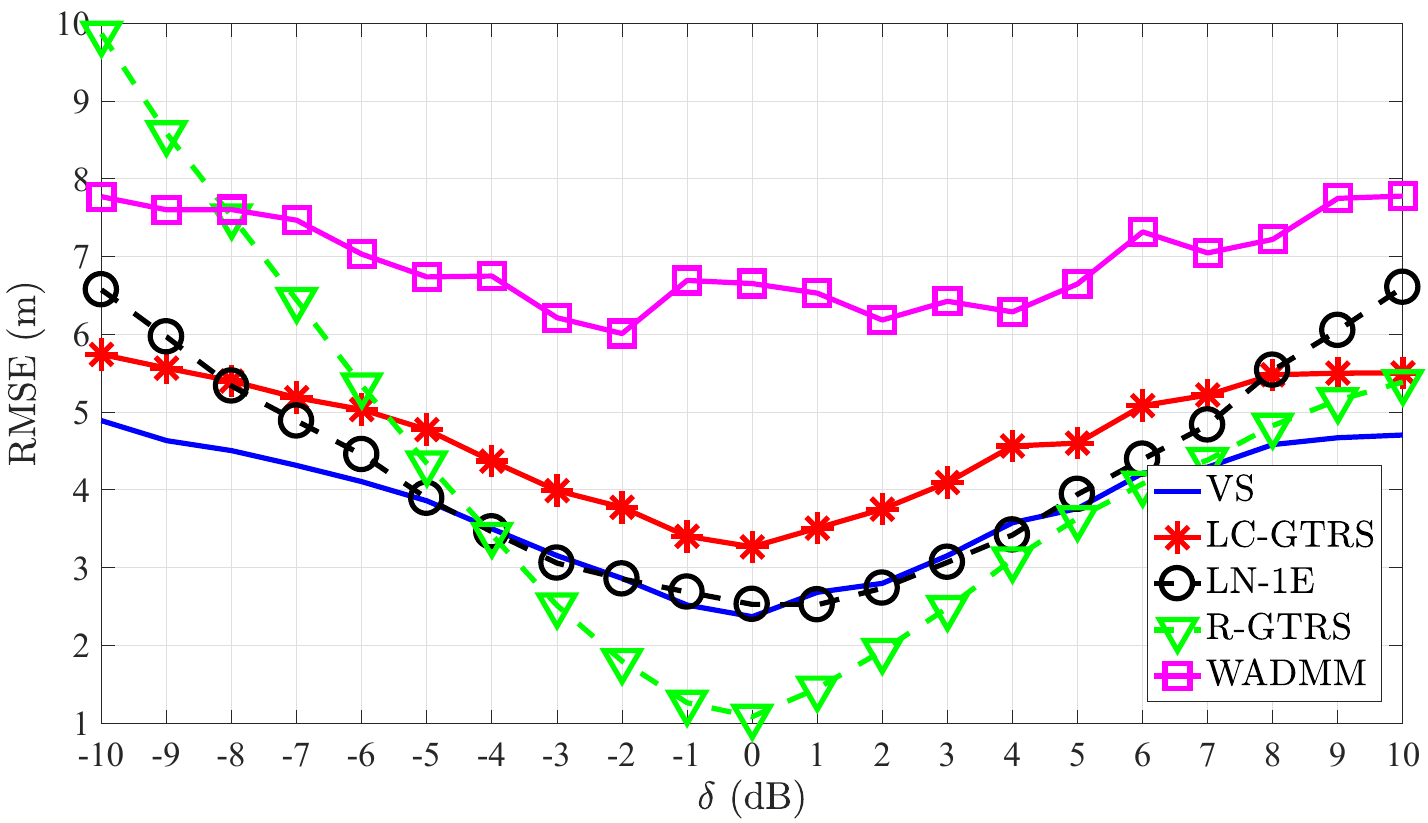}
\caption{$N = 7$ and $\sigma = 3$~(dB)}
\label{fig:RMSE_vs_delta_coord_N7_sigma3}
\end{subfigure}
\end{center}
\vspace*{0mm}
\caption{RMSE~(m) versus $\delta$~(dB) in a coordinated attack scenario.}
\label{fig:RMSE_vs_delta_coord}
\end{figure}

Fig.~\ref{fig:RMSE_vs_delta_coord} illustrates RMSE~(m) versus $\delta$~(dB) comparison of the considered approaches in a coordinated attack scenario with two malicious anchors for different values of $N$ and $\sigma$~(dB). Similar as in the uncoordinated scenario, Fig.~\ref{fig:RMSE_vs_delta_coord} exhibits superior performance of the proposed solution for medium-to-high $\abs{\delta}$ (dB), being R-GTRS and LN-1E its only competitors for low $\abs{\delta}$~(dB). This result suggests that these two methods might be somewhat more robust to noise, since low $\abs{\delta}$~(dB) can be seen as noise corruption. Still, the gain that the new method achieves for high $\abs{\delta}$ (dB) clearly compensates this behavior.

\begin{figure}
\begin{center}
\begin{subfigure}{.425\textwidth}
\hspace*{0mm}\includegraphics[width=\textwidth]{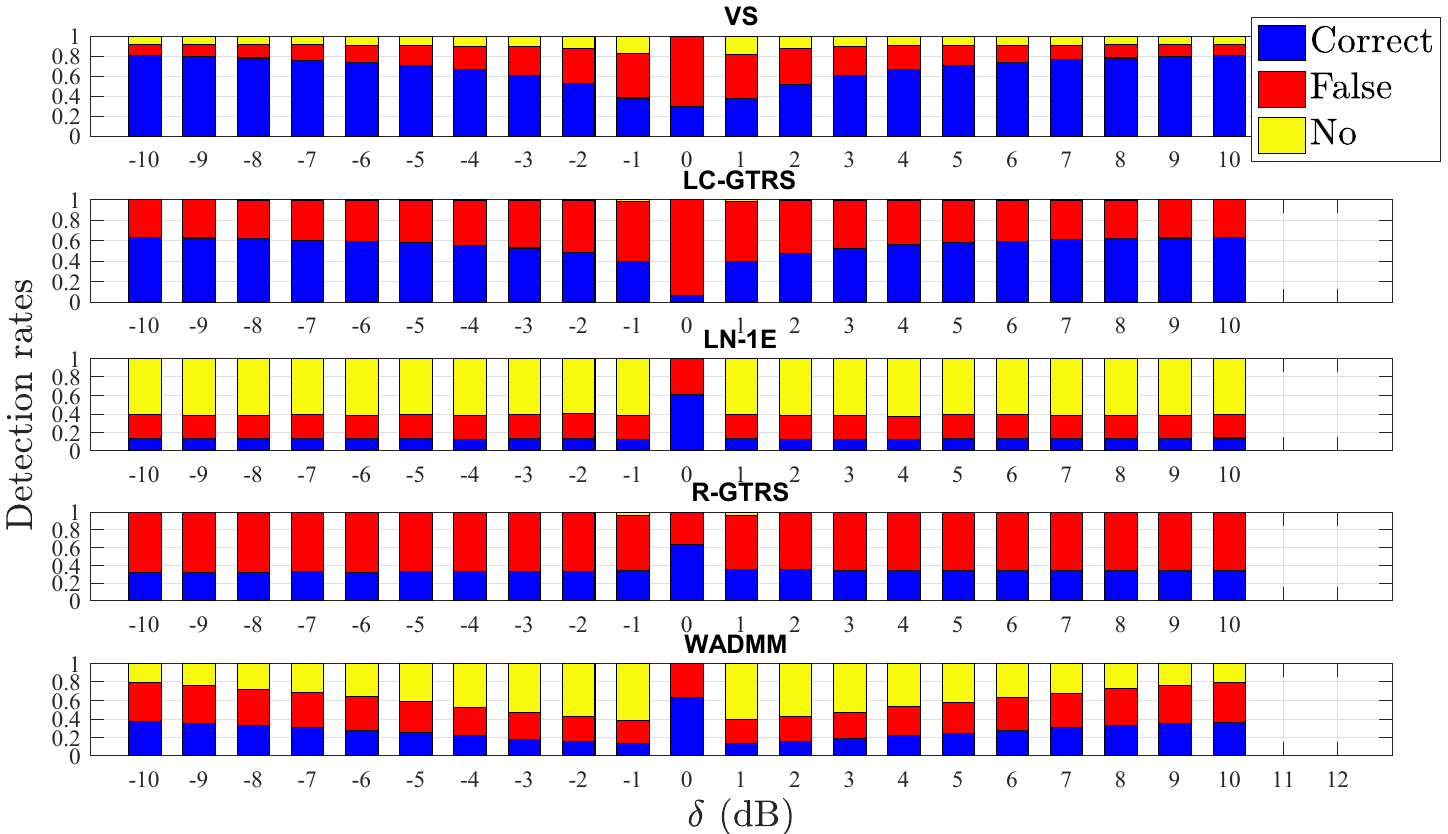}
\caption{$N = 6$ and $\sigma = 1$~(dB)}
\label{fig:DR_vs_delta_coord_N6_sigma1}
\end{subfigure}
\vspace*{0mm}
\begin{subfigure}{.425\textwidth}
\hspace*{-3mm}\includegraphics[width=\textwidth]{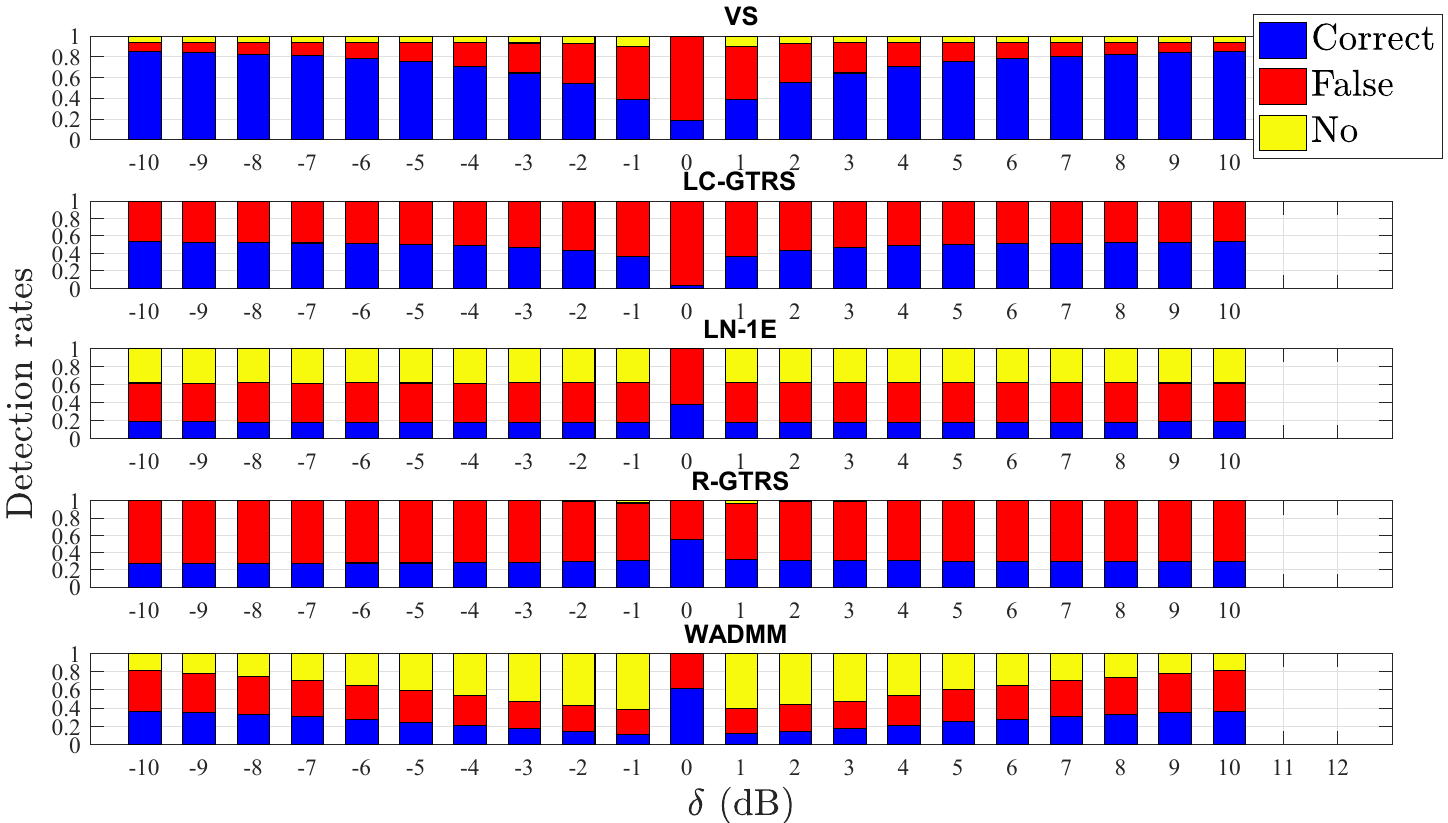}
\caption{$N = 7$ and $\sigma = 1$~(dB)}
\label{fig:DR_vs_delta_coord_N7_sigma1}
\end{subfigure}
\begin{subfigure}{.425\textwidth}
\hspace*{-3mm}\includegraphics[width=\textwidth]{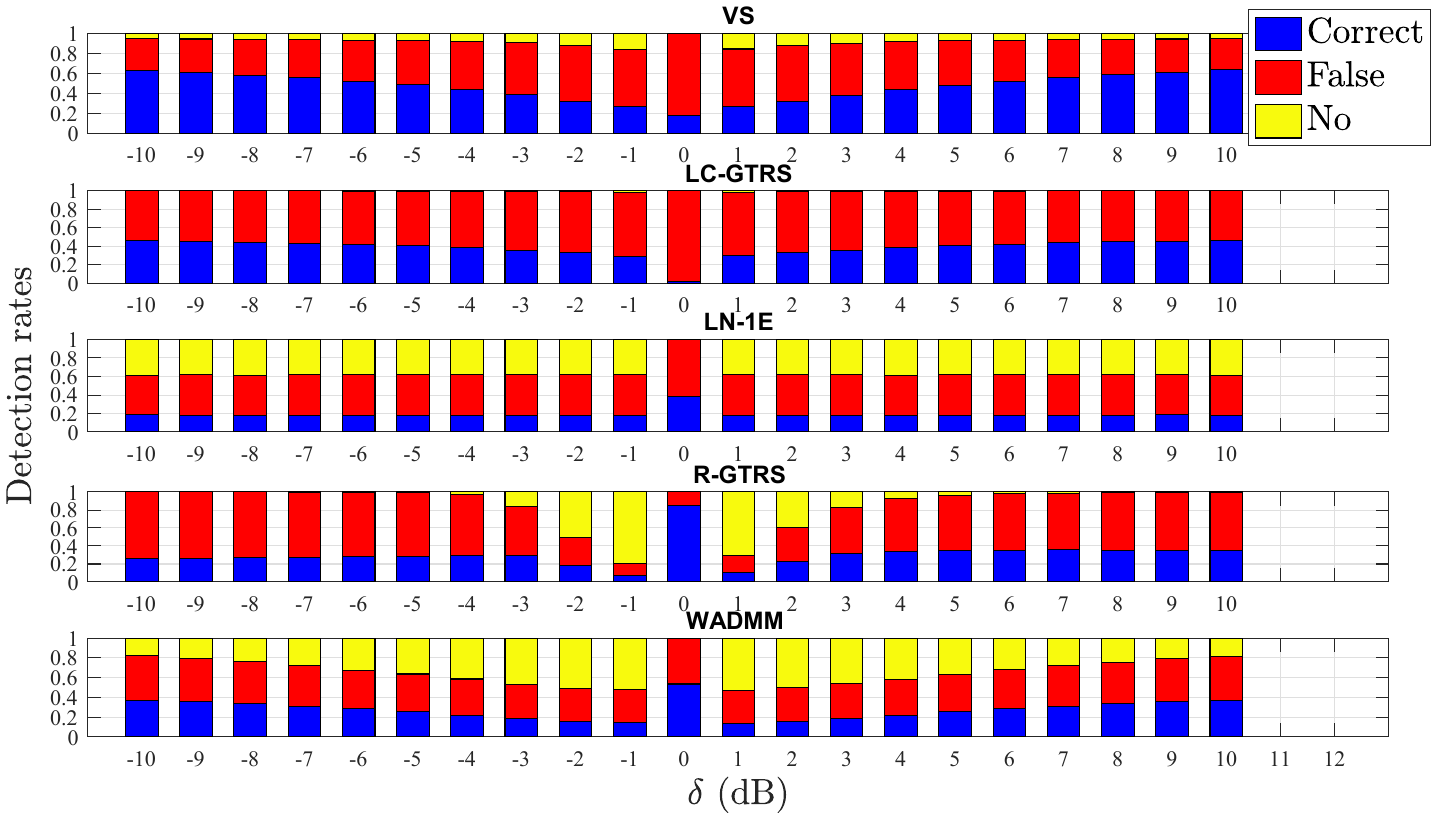}
\caption{$N = 7$ and $\sigma = 3$~(dB)}
\label{fig:DR_vs_delta_coord_N7_sigma3}
\end{subfigure}
\end{center}
\vspace*{0mm}
\caption{Detection rates versus $\delta$~(dB) in a coordinated attack scenario.}
\label{fig:DR_vs_delta_coord}
\end{figure}

Fig.~\ref{fig:DR_vs_delta_coord} illustrates different detection rates versus $\delta$~(dB) comparison of the considered approaches in a coordinated attack scenario with two malicious anchors for different values of $N$ and $\sigma$~(dB). As foreseen, (correct) detection performance of all methods degrades with the increase of $\sigma$~(dB), since higher noise power gives attackers more room to hide their attacks within the noise, which is especially visible for the ratio $\frac{\abs{\delta}}{\sigma} \in [0, 2]$~dB. Nonetheless, the proposed scheme significantly outperforms the existing ones in terms of attacker detection in general.

\begin{figure}
\centering
\includegraphics[width=.425\textwidth]{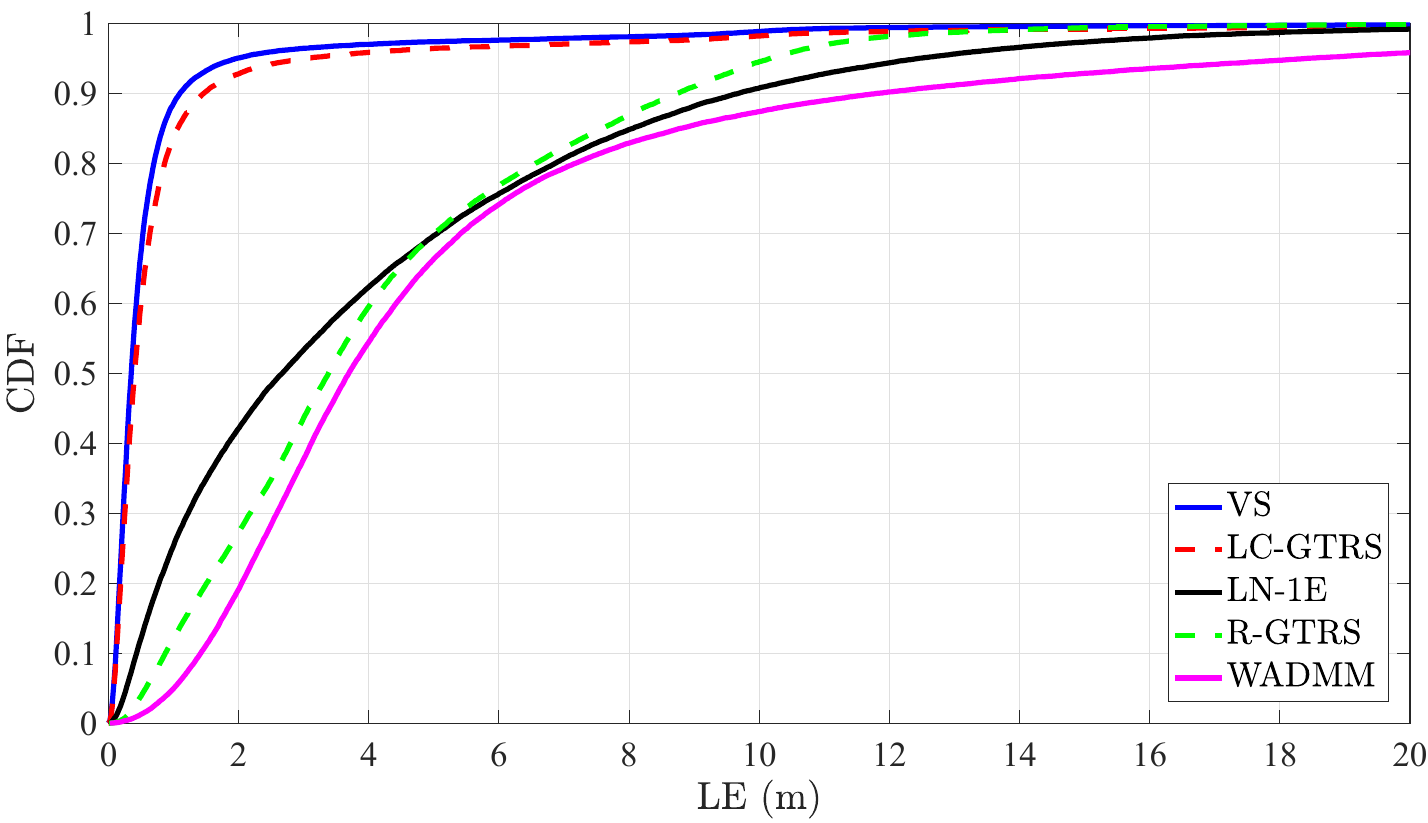}
\caption{CDF versus LE~(m) in a coordinated attack scenario, when $N = 7$ and $\sigma = 1$~(dB).}
\label{fig:CDF_vs_LE_coord}
\vspace{-3mm}
\end{figure}

Fig.~\ref{fig:CDF_vs_LE_coord} illustrates the cumulative distribution function (CDF) versus localization error (LE) in the considered coordinated scenario, when $N = 7$, $\delta = 7$~(dB) and $\sigma = 1$~(dB). Fig.~\ref{fig:CDF_vs_LE_coord} shows that the proposed scheme and LC-GTRS have almost identical performance which is clearly superior over the remaining existing ones. For instance, the former two solutions have a median of $\text{LE} \approx 0.4$~(m), whereas the best of the remaining ones exhibits a median of $\text{LE} \approx 3$~(m).

\subsection{Performance Analysis via Experimental Measurements}
\label{VOR_results}

\begin{figure}
\centering
\includegraphics[width=.4\textwidth]{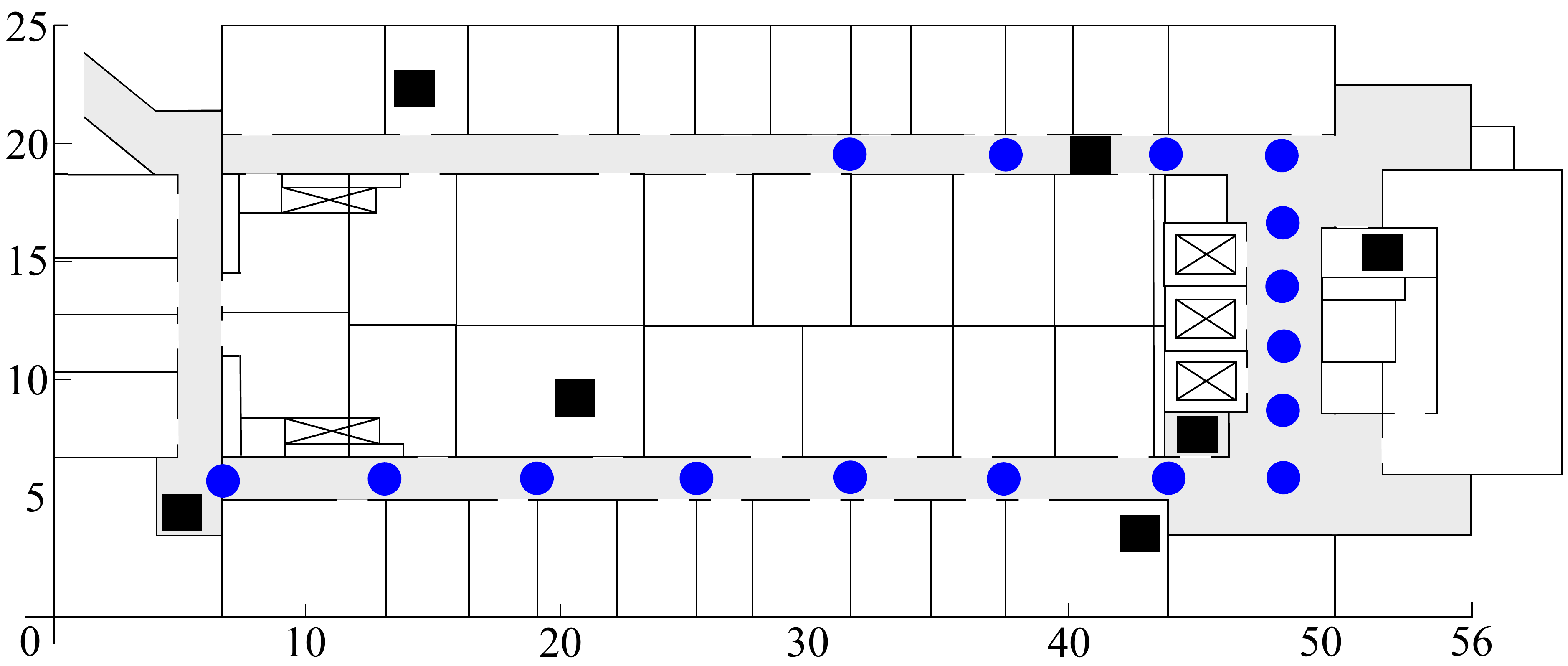}
\caption{Real-world experimental set up under consideration: blue circles and black squares respectively denote the true target and anchor locations. The data set is a courtesy of our colleagues from Computer Science department at Rutgers University~\cite{VOR}.}
\label{fig:VOR_scenario}
\vspace{-3mm}
\end{figure}

This section validates the performance of the considered algorithms in terms of localization accuracy and success of attacker detection through experimental measurements. The considered experimental configuration is illustrated in Fig.~\ref{fig:VOR_scenario}. In all experiments, $K=50$ was consider to localize a single target at a time, in the presence of two malicious anchors. Moreover, two randomly chosen anchors in each run out of $N_A=100$ runs are considered malicious.

\subsubsection{Uncoordinated Attacks}
\label{subsubsec:VOR_uncoord}

\begin{figure}
\centering
\includegraphics[width=.425\textwidth]{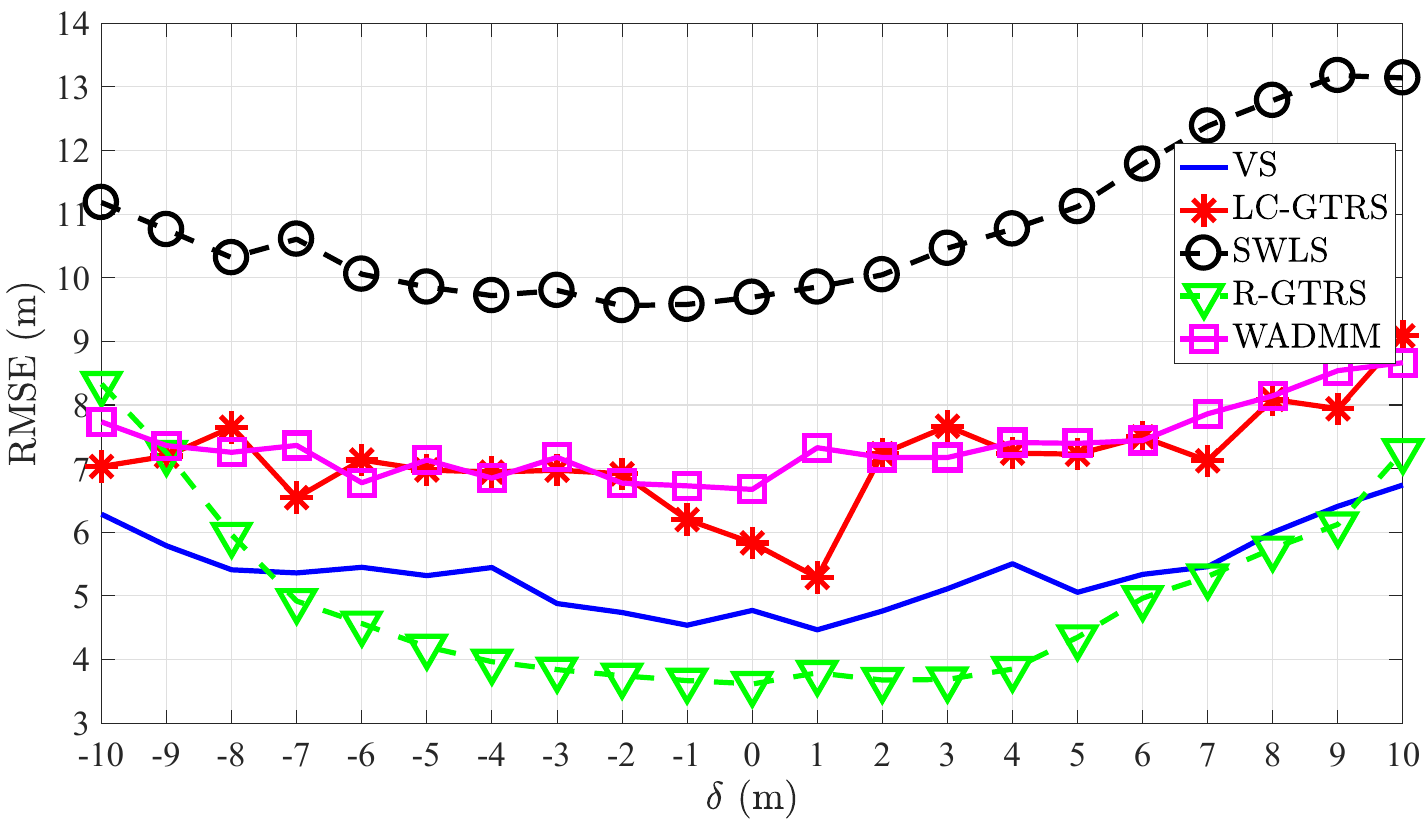}
\caption{RMSE~(m) versus $\delta$~(dB) in the considered experimental uncoordinated attack scenario, for $N_A = 100$.}
\label{fig:VOR_RMSE_vs_delta_uncoord}
\vspace{-3mm}
\end{figure}

Fig.~\ref{fig:VOR_RMSE_vs_delta_uncoord} illustrates RMSE~(m) versus $\delta$~(dB) comparison of the considered approaches in the experimental uncoordinated attack scenario with two malicious anchors. The figure exhibits that only R-GTRS performs better than the proposed scheme for low-to-medium $\delta$~(dB), whereas for high attack intensities (e.g., $\abs{\delta} \geq 7$), both methods exhibit similar performance. Nevertheless, it is worth remembering that R-GTRS requires additional knowledge about the maximum attack intensity, which is unlikely to be perfectly available beforehand in practice.

\begin{figure}
\centering
\includegraphics[width=.425\textwidth]{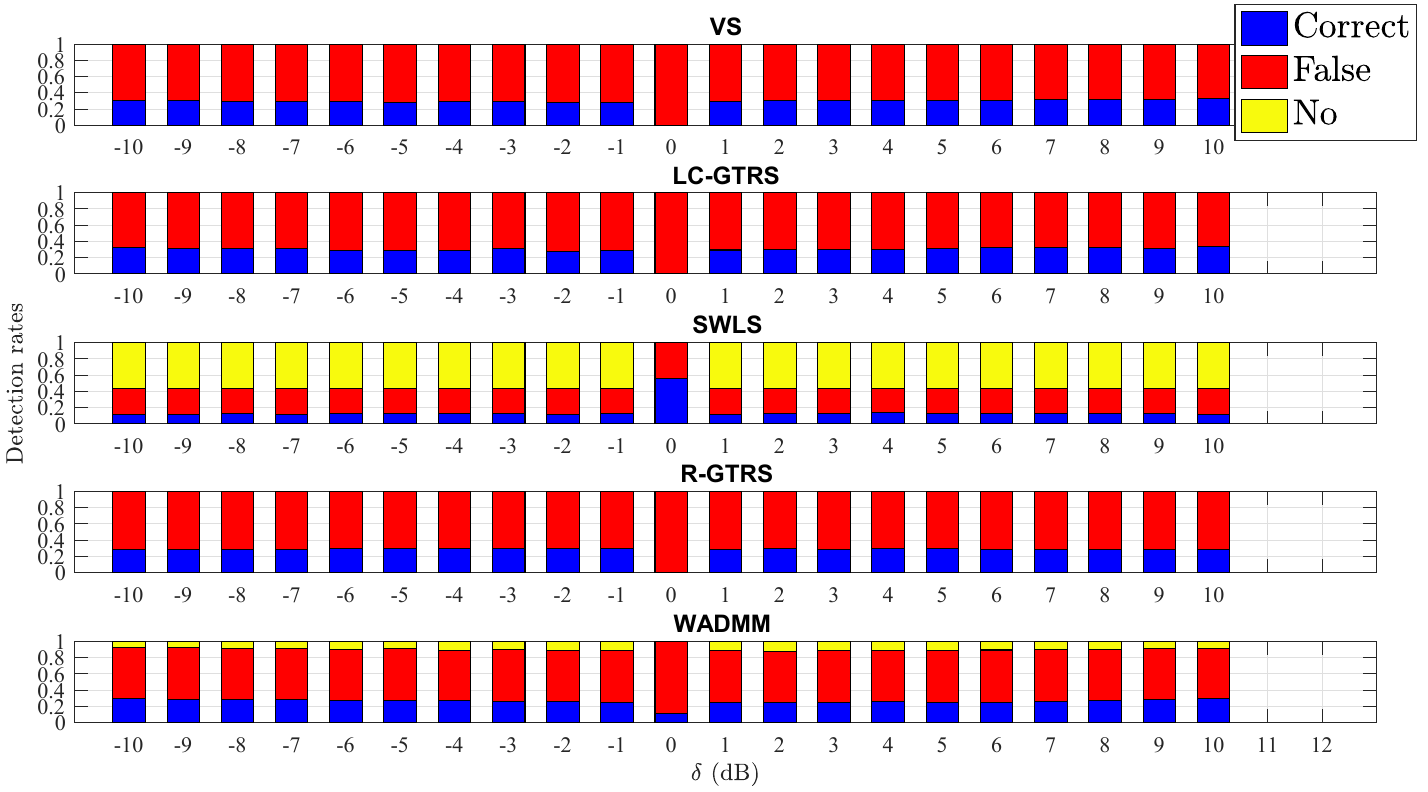}
\caption{Detection rates versus $\delta$~(dB) in the considered experimental uncoordinated attack scenario, for $N_A = 100$.}
\label{fig:VOR_DR_vs_delta_uncoord}
\vspace{-3mm}
\end{figure}

Fig.~\ref{fig:VOR_DR_vs_delta_uncoord} illustrates detection rates versus $\delta$~(dB) comparison of the considered approaches in the experimental uncoordinated attack scenario with two malicious anchors. Fig.~\ref{fig:VOR_DR_vs_delta_uncoord} shows that the achieved detection rates are significantly lower than the ones from the simulations, but this is expected given various obstacles, anisotropies and other phenomena that were not considered in simulations. Even so, the results indicate that the proposed detection scheme works well in practice, which is corroborated by the fact that its results match the best existing ones, LC-GTRS in this case, and go up to $\approx 34~\%$.

\subsubsection{Coordinated Attacks}
\label{subsubsec:VOR_coord}

\begin{figure}
\centering
\includegraphics[width=.425\textwidth]{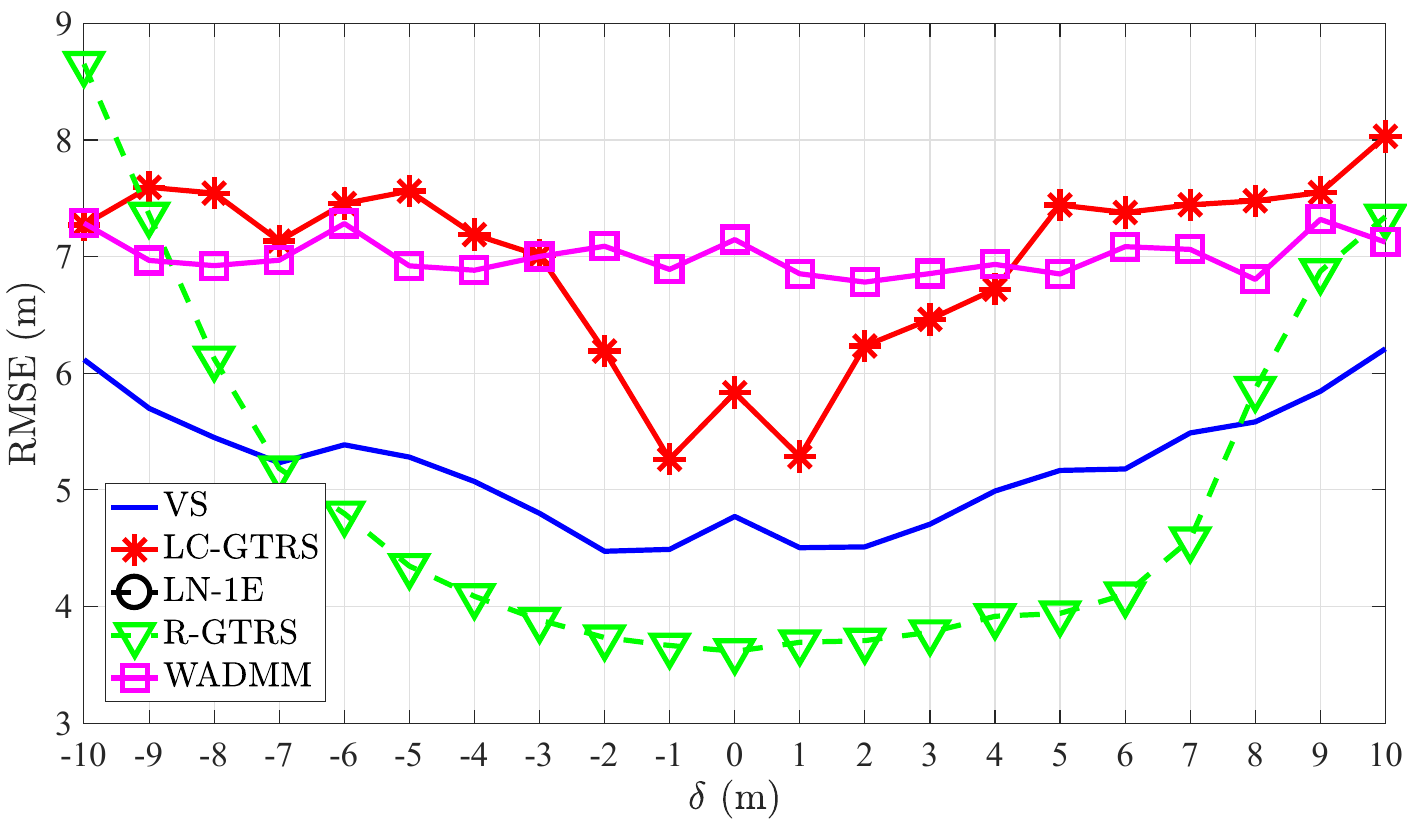}
\caption{RMSE~(m) versus $\delta$~(dB) in the considered experimental coordinated attack scenario, for $N_A = 100$.}
\label{fig:VOR_RMSE_vs_delta_coord}
\vspace{-3mm}
\end{figure}

Fig.~\ref{fig:VOR_RMSE_vs_delta_coord} illustrates RMSE~(m) versus $\delta$~(dB) comparison of the considered approaches in the experimental coordinated attack scenario with two malicious anchors. Similar as in the uncoordinated case, only R-GTRS performs better than the proposed scheme for low-to-medium $\delta$~(dB), whereas for high attack intensities (e.g., $\abs{\delta} \geq 7$), the proposed approach exhibits the best performance.

\begin{figure}
\centering
\includegraphics[width=.425\textwidth]{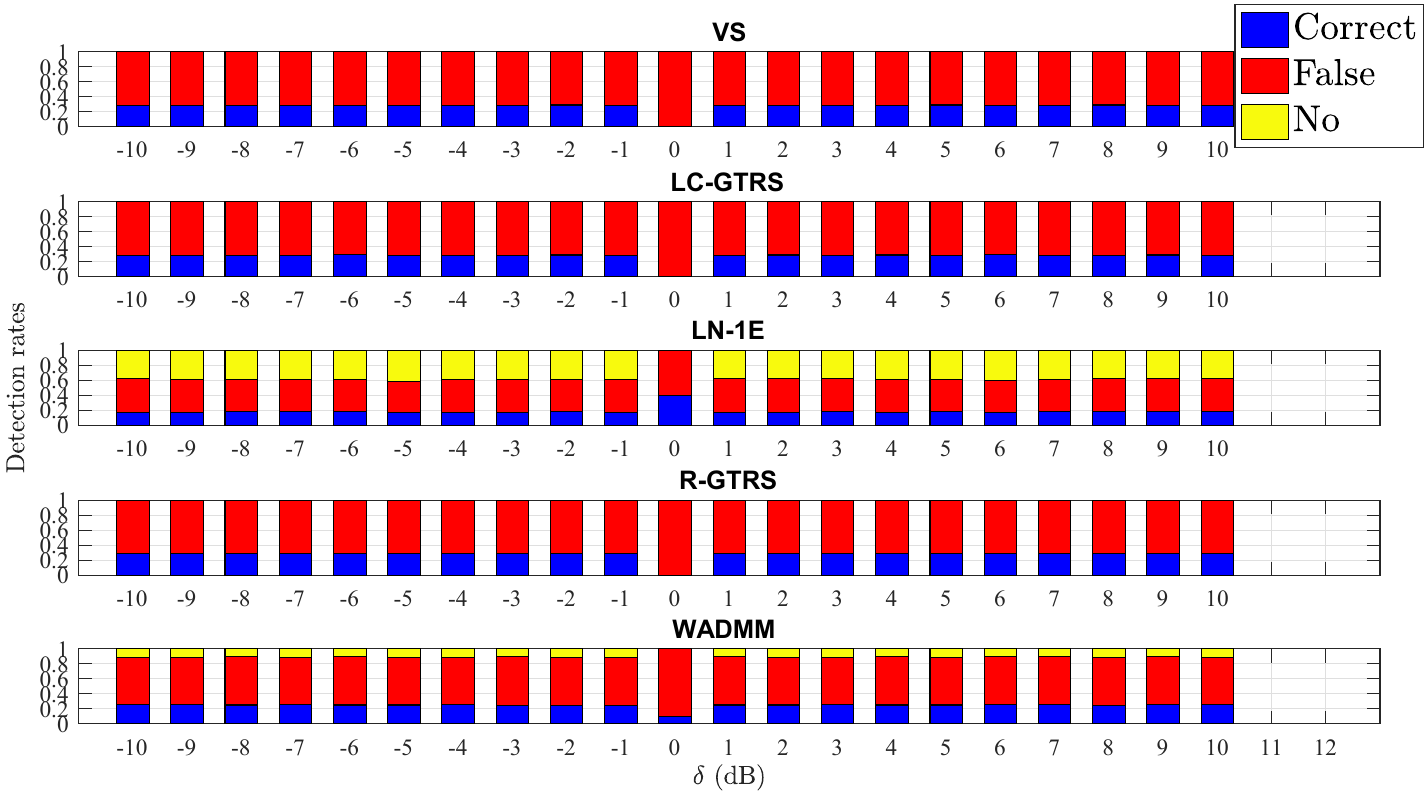}
\caption{Detection rates versus $\delta$~(dB) in the considered experimental coordinated attack scenario, for $N_A = 100$.}
\label{fig:VOR_DR_vs_delta_coord}
\vspace{-3mm}
\end{figure}

Fig.~\ref{fig:VOR_DR_vs_delta_coord} illustrates detection rates versus $\delta$~(dB) comparison of the considered approaches in the experimental coordinated attack scenario with two malicious anchors. Naturally, the figure shows a drop in the detection rates when compared with the ones in the simulations. Again, the proposed detection scheme is corroborated in practice, matching the performance of the best existing one, and goes up to $\approx 30~\%$.

\subsection{Discussion}
\label{subsec:discus}

While the proposed solution outperforms existing methods in localization and detection for medium-to-high attack intensity, some limitations and parameter choices are worth discussing. Certain parameters are fixed, such as the number of points of interest set to $N-1$ for localization and the confidence interval for attacker detection set within one estimated noise standard deviation. The rationale for using  $N-1$ points is that not all $N$ intersection points are necessary, as some may be affected by attacks or excessive noise. For smaller $N$, trimming unnecessary points improves accuracy. The confidence interval choice balances false detection (if too narrow) and missed detection (if too wide), with one standard deviation offering a reasonable trade-off. Although these parameters could be fine-tuned for different scenarios, numerical results indicate that the chosen values work well within the considered settings.

The proposed approach assigns beliefs (votes) to each intersection point, but under high noise or unfavorable attack coordination, it may mistakenly remove a genuine point, potentially impacting localization accuracy. However, numerical results confirm that such occurrences are infrequent and do not significantly degrade performance. While the final detection output is a binary decision (genuine or malicious anchor), this classification is solely for detection purposes and does not influence localization refinement. Like existing methods, the approach requires at least $50~\%$ of anchors to be non-compromised, particularly in coordinated attacks, as a malicious majority would easily manipulate localization. A unique aspect of the method is its ability to generate points of interest even when anchor circles do not intersect. In extreme cases, all points might need to be fabricated, leading to inaccurate clusters. However, this scenario poses a challenge for all existing methods, as they inherently rely on intersection points for localization.

\begin{figure}
\begin{center}
\begin{subfigure}{.425\textwidth}
\hspace*{0mm}\includegraphics[width=\textwidth]{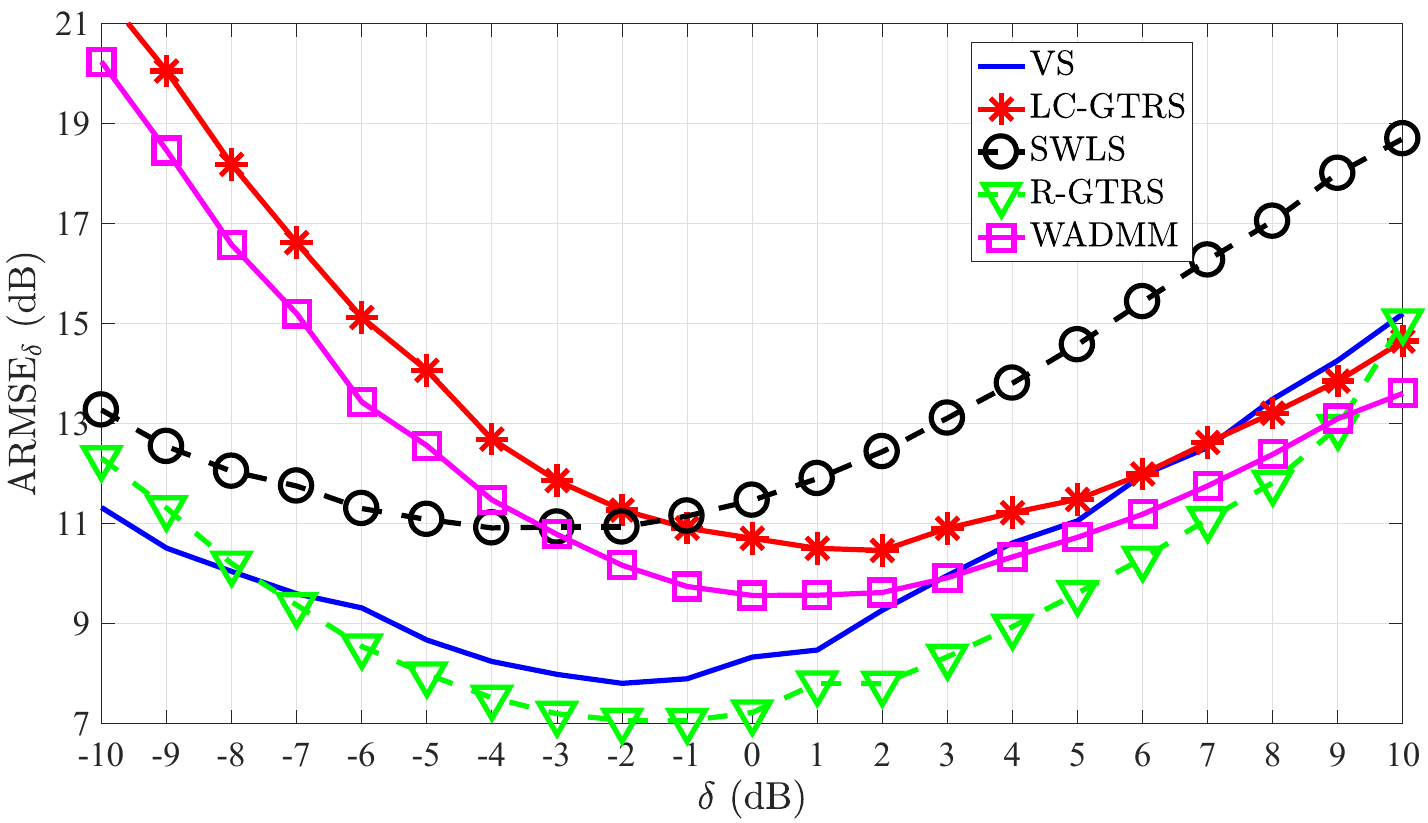}
\caption{Uncoordinated attacks}
\label{fig:VOR_ARMSE_vs_delta_uncoord}
\end{subfigure}
\vspace*{0mm}
\begin{subfigure}{.425\textwidth}
\hspace*{-3mm}\includegraphics[width=\textwidth]{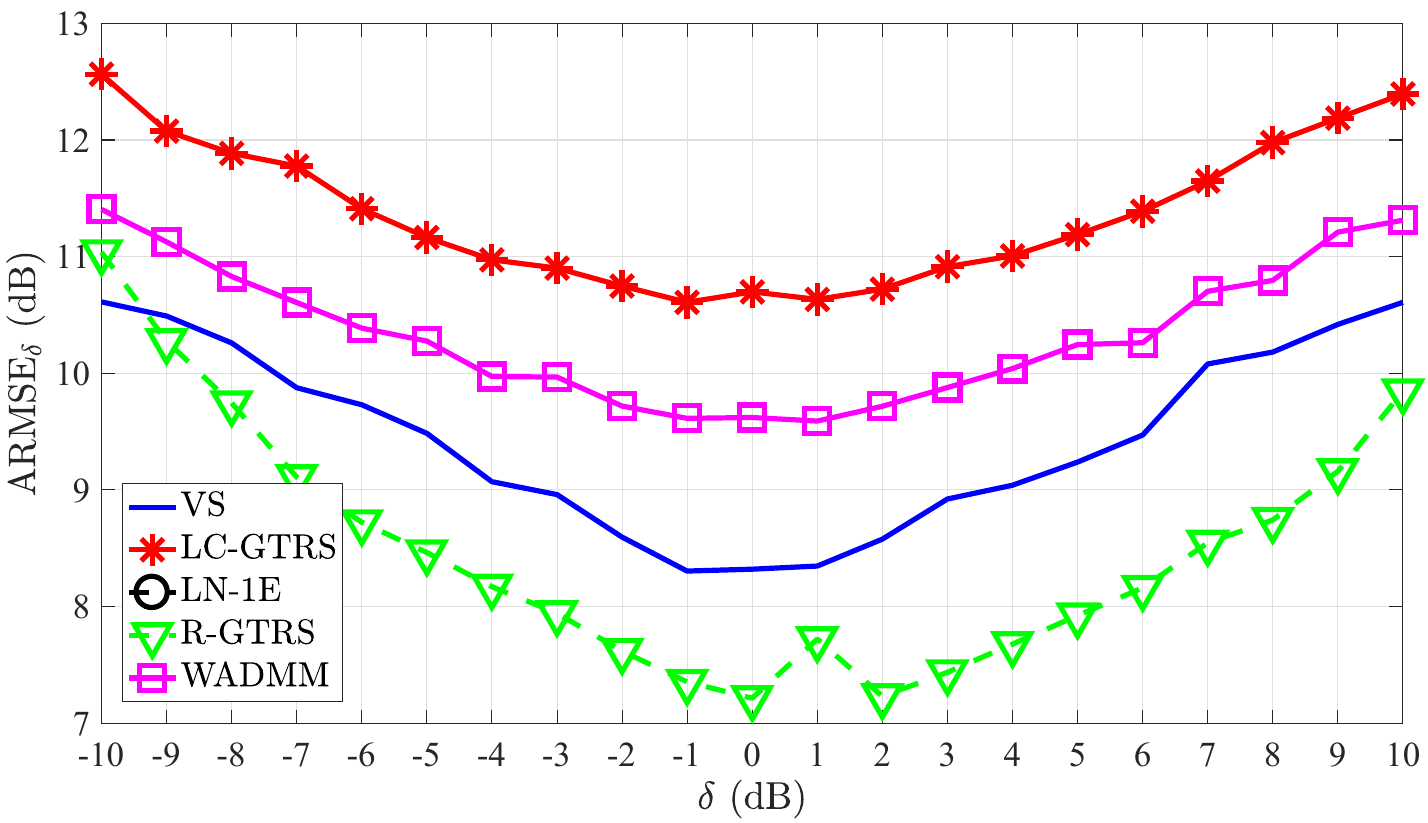}
\caption{Coordinated attacks}
\label{fig:VOR_ARMSE_vs_delta_coord}
\end{subfigure}
\end{center}
\vspace*{0mm}
\caption{$\text{ARMSE}_{\delta}$ versus $\delta$~(dB) in the considered experimental scenario, for $N_A = 100$.}
\label{fig:VOR_ARMSE_vs_delta}
\end{figure}

Although not all considered methods estimate the attack intensity directly in their detection procedure, one can obtain this estimate as explained here in~\eqref{eq:attack_estimate}. Hence, Fig.~\ref{fig:VOR_ARMSE_vs_delta} illustrates the average RMSE of $\boldsymbol{\delta} = [\delta_i]^T$ ($\text{ARMSE}_{\delta}$) in~(dB) versus $\delta$~(dB) in the considered experimental uncoordinated attack scenario, Fig.\ref{fig:VOR_ARMSE_vs_delta_uncoord}, and the considered experimental coordinated attack scenario, Fig.\ref{fig:VOR_ARMSE_vs_delta_coord}, for $N_A = 100$. The figure shows that none of the considered approaches accomplishes an error below $7$~(dB), which is due to the complexity of the scenario. Still, the proposed scheme exhibits a fairly good, but in general not the best, estimation performance, indicating that this estimation is not a prerequisite for high detection rate, which is somewhat intuitive.



\section{Conclusions}
\label{sec:conclusions}

This work advances target localization in randomly deployed WSNs under both uncoordinated and coordinated spoofing attacks by introducing a novel geometric approach for accurate localization and attacker detection. The method estimates the target's location using intersection points of anchor pairs, applying a VS and WCM, followed by attacker detection via confidence intervals. Unlike traditional methods, it employs soft detection decisions, assigning beliefs (votes) instead of binary classifications. These votes are converted into probabilities and used as WCM weights for refined localization. Potential attack intensities are estimated using an ML criterion, with final detection based on confidence intervals. Performance evaluation in terms of localization accuracy, detection rate, and computational complexity shows that the method achieves near-perfect detection when the attack intensity-to-noise ratio is sufficiently high. Although detection rate decreases as this ratio lowers, the proposed approach consistently outperforms existing methods in both detection performance and localization accuracy, reducing localization error by at least $30~\%$ across various scenarios. Additionally, the VS-based solution operates efficiently, with a runtime of just a few milliseconds.



\bibliographystyle{IEEEtran}
\bibliography{References}



\vspace*{5mm}
\begin{wrapfigure}{l}{25mm} 
\includegraphics[width=1in,height=1.25in,clip,keepaspectratio]{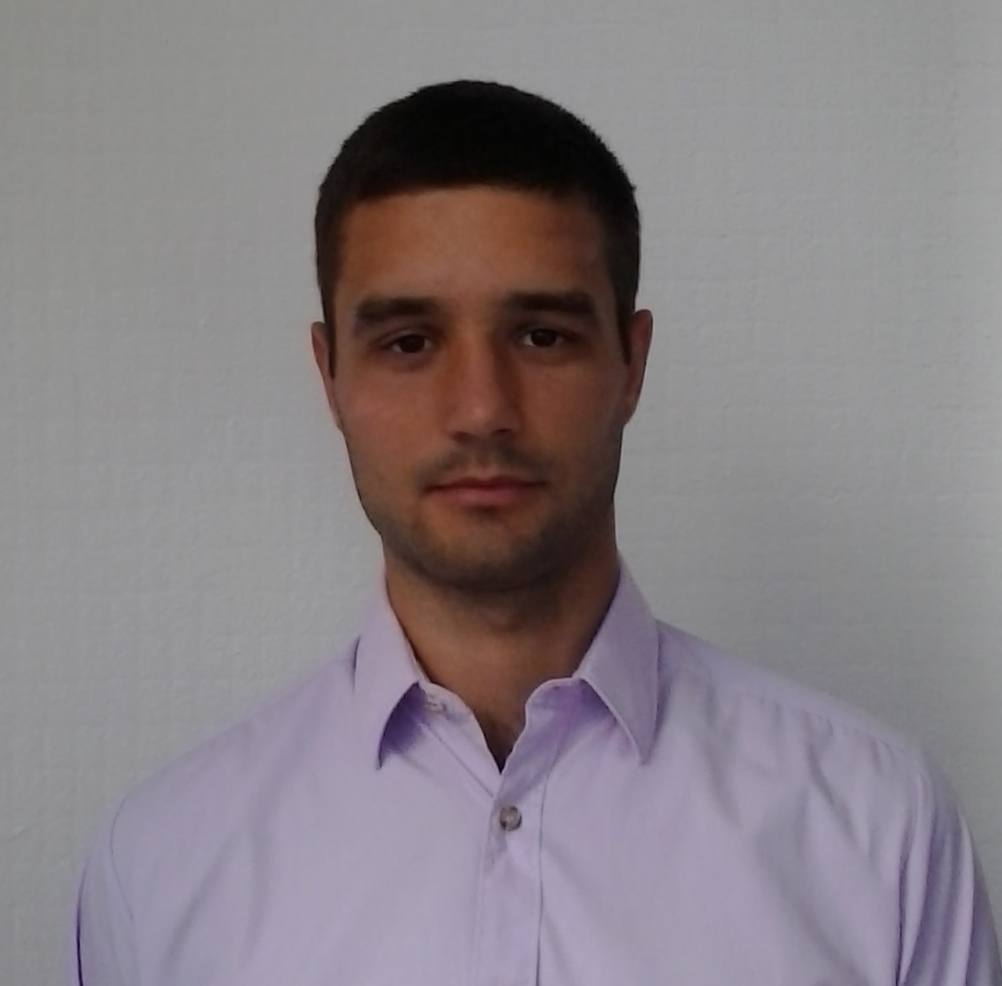}
\end{wrapfigure}\par
\textbf{Slavisa Tomic} received the M.S. degree in traffic engineering according to the postal traffic and telecommunications study program from University of Novi Sad, Serbia, in 2010, and the PhD degree in electrical and computer engineering from University Nova of Lisbon, Portugal, in 2017. He is currently an Associate Professor at the Universidade Lus\'{o}fona, Lisbon, Portugal. He is one of the winners of the 4th edition of Scientific Employment Stimulus (CEEC Individual 2021) funded by Fundação para a Ciência e a Tecnologia. According to the methodology proposed by Stanford University, he was among the most influential researchers in the world between 2019 and 2023 when he joined the list of top $2~\%$ of scientists whose work is most cited by other colleagues in the field of Information and Communication Technologies, sub-area Networks and Telecommunications. His research interests include target localization in wireless sensor networks, and non-linear and convex optimization.\par

\vspace*{5mm}
\begin{wrapfigure}{l}{25mm} 
\includegraphics[width=1in,height=1.25in,clip,keepaspectratio]{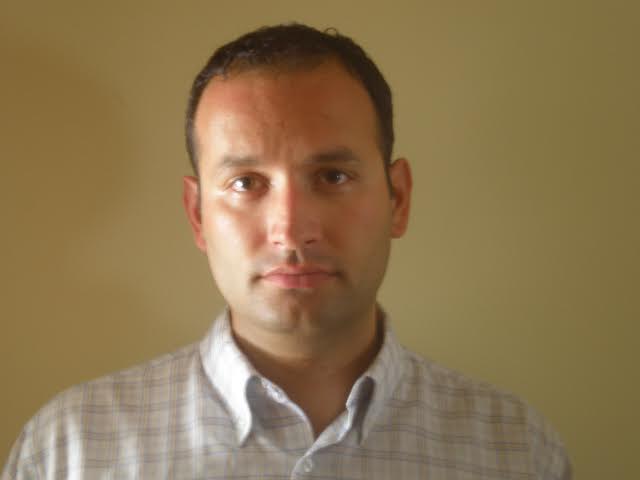}
\end{wrapfigure}\par
\textbf{Marko Beko} was born in Belgrade, Serbia, in November 1977. He received the Ph.D. degree in electrical and computer engineering from the Instituto Superior T\'{e}cnico (IST), Universidade de Lisboa, Portugal, in 2008. He received the title of a ``Professor with Habilitation'' of electrical and computer engineering from the Universidade Nova de Lisboa, Lisbon, in 2018. His current research interests lie in the area of signal processing for wireless communications. He serves as an Associate Editor for the IEEE Open Journal of the Communications Society. He is also a member of the Editorial Board of IEEE Open Journal of Vehicular Technology. He is the winner of the 2008 IBM Portugal Scientific Award. According to the methodology proposed by Stanford University, he was among the most influential researchers in the world between 2019 and 2023 when he joined the list of top $2~\%$ of scientists whose work is most cited by other colleagues in the field of Information and Communication Technologies, sub-area Networks and Telecommunications. He is one of the founders of Koala Tech.\par

\vspace*{5mm}
\begin{wrapfigure}{l}{25mm} 
\includegraphics[width=1in,height=1.25in,clip,keepaspectratio]{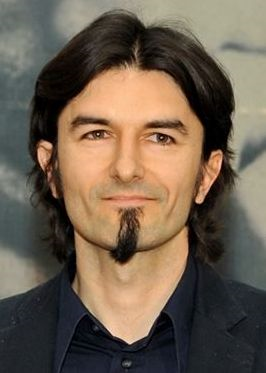}
\end{wrapfigure}\par
\textbf{Dejan Vukobratovic} received the Ph.D. degree in electrical engineering from the University of Novi Sad, Serbia, in 2008. From 2009 to 2010, he was a Marie Curie Intra-European Fellow with the University of Strathclyde, Glasgow, U.K. Since 2019, he has been a Full Professor with the Department of Power, Electronics and Communication Engineering, University of Novi Sad. His research interests include wireless communication systems and the Internet of Things.\par

\vspace*{5mm}
\begin{wrapfigure}{l}{25mm} 
\includegraphics[width=1in,height=1.25in,clip,keepaspectratio]{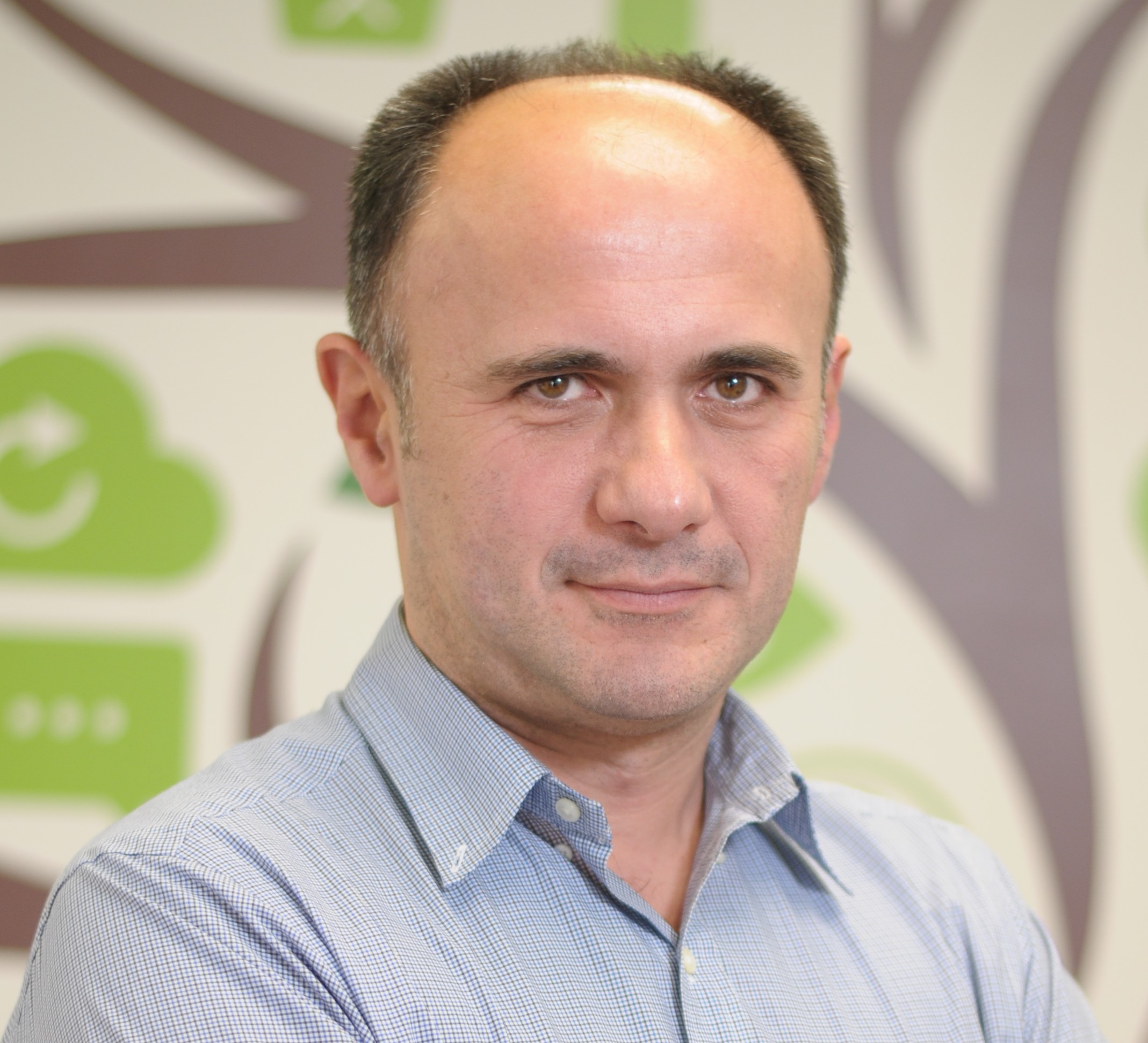}
\end{wrapfigure}\par
\textbf{Srdjan Krco} received his PhD in electrical engineering in 2005 for the work on remote health monitoring systems. He was with Ericsson for over 10 years, working on the development of 3G networks, M2M and IoT technologies. Since 2010, he manages DunavNET, a company he co-founded. DunavNET designs turnkey solutions based on IoT and ML/AI technologies for agriculture, manufacturing, and cities.  He is active in international collaborative research and innovation projects (e.g., H2020, Horizon Europe) and is teaching IoT, digital transformation, and entrepreneurship at several universities in Europe and Africa. He has a dozen patents in his name and has published a large number of papers in international journals and conferences. He received the Innovation Engineer of the Year award in 2007 from the Irish Institute of Engineers and is being awarded Microsoft MVP annual award since 2017. He is a member of the Management Board of AIOTI (www.aioti.eu).\par


\end{document}